# Tackling the challenge of a huge materials science search space with quantum-inspired annealing

*Kan Hatakeyama-Sato[a], Takahiro Kashikawa[b], Koichi Kimura[b], and Kenichi Oyaizu[a]\**

a) Department of Applied Chemistry Waseda University Tokyo 169-8555, Japan
b) Fujitsu Laboratories Ltd., Kanagawa 211-8588, Japan

E-mail: oyaizu@waseda.jp

Abstract

Efficient screening of chemicals is essential for exploring new materials. However, the search space is astronomically large, making calculations with conventional computers infeasible. For example, an *N*-component system of organic molecules generates $>10^{60N}$ candidates. Here, a quantum-inspired annealing machine is used to tackle the challenge of the large search space. The prototype system extracts candidate chemicals and their composites with desirable parameters, such as melting temperature and ionic conductivity. The system can be at least $10^4$–$10^7$ times faster than conventional approaches. Such exponential acceleration is critical for exploring the enormous search space in virtual screening.

# 1. Introduction

Finding new functional materials with improved performance is critical, especially in developing next-generation batteries, solar cells, and other energy-related devices.[1] Materials informatics has been attracting attention because of its potential to accelerate research and development.[1, 2] Machine learning, which can handle the big data from materials science and identify statistical trends, is a critical technology in this field.[1-3] Trained models can predict the performance ($y$) of diverse materials from their structures ($x$).[1, 2]

The predicted values ($y_{pred}$) from a trained model ($f_{ML}$) have become increasingly reliable. Various parameters, including basic molecular properties (e.g., mechanical strength and melting point),[4] conductivity,[3] photoconversion efficiency,[5] and synthetic yields,[6] have been predicted successfully. If an appropriate training database and model are prepared, the predictions will become as accurate as those from human researchers and molecular simulations.[3] Another major advantage of machine learning is its high prediction speed (e.g., $10^{-3}$ s per condition),[1] allowing researchers to screen chemicals with a trained model much faster than with real experiments and first-principles calculations.

A key challenge in materials informatics is the astronomically large search space for candidate materials. Even for small organic molecules, there are more than $10^{60}$ candidates.[7] Current virtual compound screening using a machine learning function, $y_{pred} = f_{ML}(x)$, may not finish before the end of the universe; the required calculation time of $10^{60} \times 10^{-3}$ s is longer than the estimated life of the universe, $10^{19}$ s [8] (Figure 1). Furthermore, the search space increases exponentially when a composite system of multiple chemicals is considered ($10^{60N}$ candidate composites for an $N$-component system), and most conventional materials and devices consist of multiple components.[1, 3, 5]

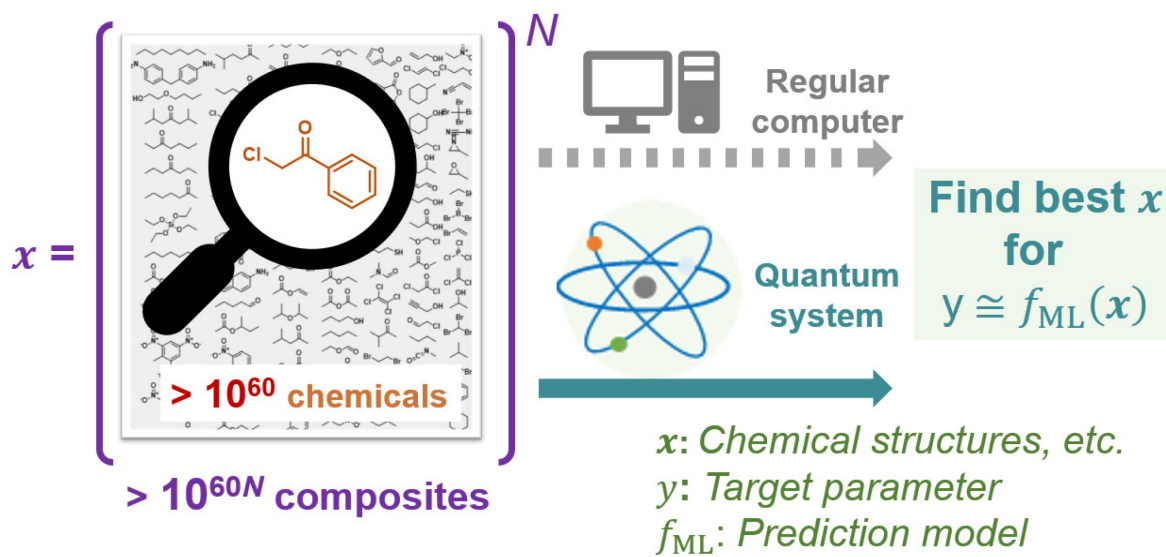

**Figure 1.** Exploration of $10^{60}$ chemicals and their composites by conventional computers and quantum systems.

Various approaches have been proposed to solve the massive search space problem, such as defining an inverse function $x = f_{ML}^{-1}(y)$, which predicts structures or experimental conditions from a target parameter $y$.[9-12] However, the function is intrinsically hard to define because of the uniqueness of solution problem.[1, 10, 12] Furthermore, the preparation of chemical structures itself is still a huge challenge even with cutting-edge deep learning techniques due to the current limitations of model complexity and computing power.[13] Other molecule generation techniques, such as Bayesian approaches,[14] also tend to have similar limitations. Most studies have focused on single-component systems and the methodology for multiple-component systems has not been developed satisfactorily.[10, 11, 14]

A practical approach to exploring materials is to narrow the search space by filtering candidates manually. Based on the knowledge of human researchers, only feasibly useful structures are selected.[3, 5] Typically, $10^2$–$10^4$ candidate chemicals would be selected to allow calculations to be finished quickly[3, 5, 9]. However, filtering based on existing knowledge may overlook hidden potential candidates. The approach also may fail when treating *N*-component systems. If there are $10^M$ chemical candidates, their available combinations will be $10^{M \times N}$, excluding their weight compositions (i.e., $N$ - 1 additional independent variables). For instance, the calculation time would be as long as $10^9 \times 10^{-3}$ s $\cong$ 300 h with $M \times N = 9$. This type of combinatorial explosion is critical because $M \geq 3$ and $N \geq 3$ are typical for conventional materials and devices.[3, 5]

Here, we introduce a quantum physics-inspired annealing machine to tackle the challenge of the huge search space (Figure 1). The main limitation of conventional approaches is the insufficient computing power compared with the search space. Quantum physics-based or -inspired computing solves the exploration problems efficiently[15-20] owing to massive parallelization that cannot be achieved by conventional computers.[15-20] To demonstrate the potential of quantum computing approaches, we constructed a prototype system to explore organic chemicals and composites displaying desirable parameters (e.g., melting point and ionic conductivity). The system was at least $10^4$–$10^7$ times faster than conventional approaches, helping to solve the combinatorial explosion problem in materials science.

## 2. Molecular exploration system framework

Quantum computers can solve some types of problems quickly, including combinatorial optimization, prime factorization, and molecular quantum processes.[16, 19] The superposition principle is useful for exploring new materials from a vast search space efficiently (Figure 2a). Gate-based quantum computers can solve various problems, although they currently have many technical problems, such as long-time coherence.[16, 19] In contrast, quantum annealing machines have already been commercialized.[21] Quantum annealing machines are used to solve the Ising model (equation 1),[15, 21, 22] which was used to explore materials in our current study. Previously, annealers have been used in limited cases to explore materials (e.g., orders of nanomaterials).[15] In this work, our prototype system was used to explore general organic materials with diverse properties.

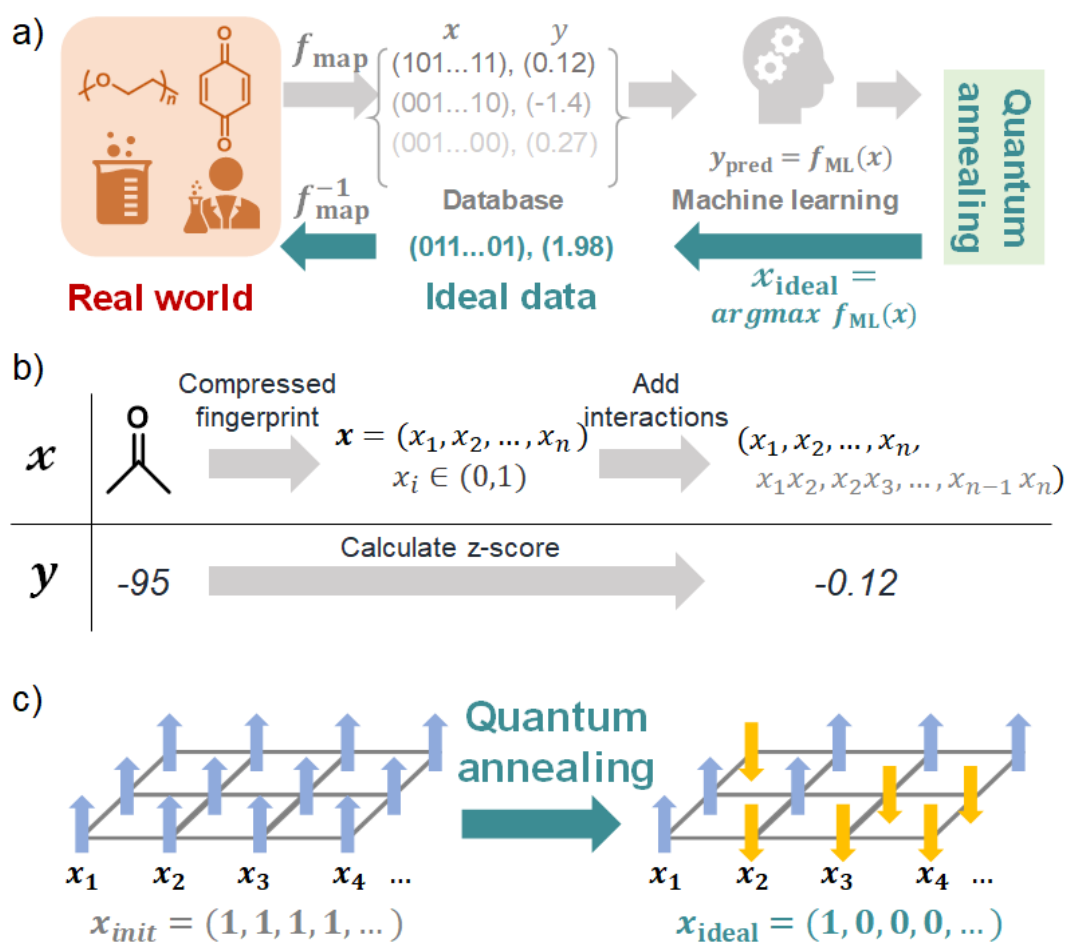

**Figure 2.** a) General scheme for exploring the best input ($x_{ideal}$) by quantum(-inspired) annealing. b) Conversion of chemical structures into binary arrays. c) Overview of quantum(-inspired) annealing.

The architecture of our material screening system is shown in Figure 2a. First, structural information was converted into binary arrays (0 or 1) using a mapping function, $f_{map}$ (Figure 2b). Binarization was needed because quantum annealers accept only binaries.[22] A machine learning function, $f_{ML}$, was prepared in the form of the Ising model. An experimental database of target materials was used to train the model.

The machine learning function was a linear model considering quadratic terms ($x_i x_j$, $i \neq j$). Annealers can solve the problem of $\boldsymbol{x_{ideal}} = argmax\ f_{ML}(\boldsymbol{x})$ (or $argmin$) promptly (around $10^{-6}$ to $10^{-0}$ s, Figure 2c).[20, 22] The obtained $\boldsymbol{x_{ideal}}$ could give $y_{pred}$, which is close to the global maximum (or minimum) of the system (equation 1).[22]

$$y_{pred} = f_{ML}(\boldsymbol{x}) = \sum_{i \neq j} J_{ij}\, x_i x_j + \sum_i h_i\, x_i \qquad (1)$$

Conventional computers may not be able to solve the problem when the dimension of $\boldsymbol{x}$ becomes higher than around 100 (i.e., $2^{100} \cong 10^{30}$-dimensional search space).[20] Finally, chemical structures were proposed as real-world information based on an inverse mapping function, $f_{map}^{-1}$.

For the annealer, we used a digital annealing unit, which was developed based on quantum annealing.[20] The machine is a digit simulator, but it can treat up to 8192 fully connected binary variables with massive parallelization.[20] In contrast, the commercialized quantum annealer D-Wave still has a strict limitation on the input variables and binary interactions; there are typically fewer than 100 available variables for achieving fully connected bits.[8, 21, 22]

## 3. Exploration with a single-component system

First, we extracted high-melting-temperature molecules with the annealer. An open experimental database of organic molecules[23] was used for machine learning (about 3000 chemicals, Table S1). The structural information was converted into 2048-dimensional molecular fingerprints $x_{FP}$.[24] The array was compressed to $n$-dimensional arrays to obtain a binary input, $x$ ($0 < n < 2048$, Figures 2b, 3a, S1, and S2). Compression was needed because $_nC_2$ interactions must be added as quadratic terms $x_i x_j$ (i.e., $_{2048}C_2 = 2{,}096{,}128$ terms with $n = 2048$ are too many for regression). After adding the terms, the statistical relationships between the input and $z$-scores of the melting temperatures ($y$) were inputted to train a linear regression model (Figure 2a and equation 1). The regression determined the coefficients $J_{ij}$ and $h_i$.

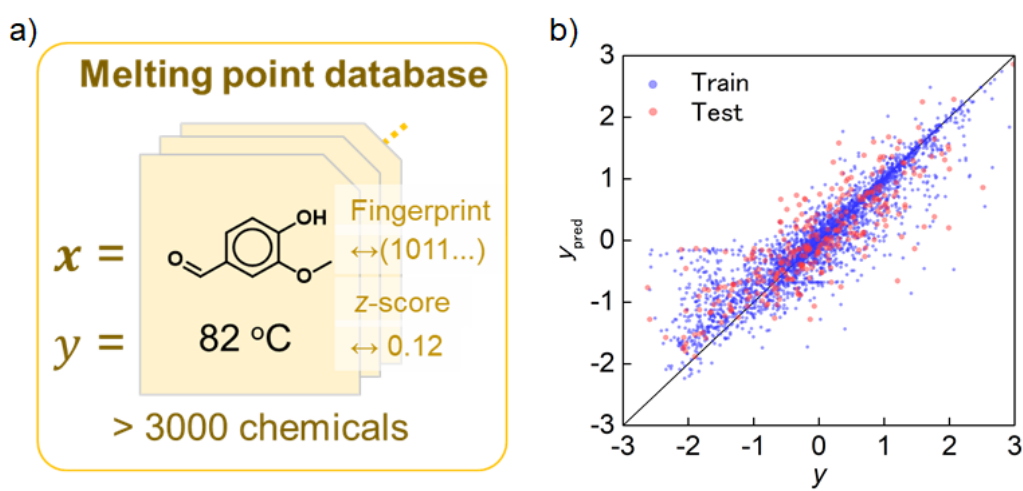

**Figure 3.** a) Scheme of preparing $x$ and $y$ for the melting point database. b) Regression result of melting point (as a $z$-score, $y$). The database was split into the training (90%) and testing (10%) datasets, for which the $R^2$ scores were 0.75 and 0.51, respectively. Ridge regression was conducted with a regularization strength $\alpha = 0.1$. The dimension of $x$ was 576.

Regularized models[25] were used for regressions. Ridge (i.e., L2 norm) models gave better prediction performances than did Lasso models (i.e., L1 norm, Figure S3). The sparsity obtained by the Lasso model may have discarded too much of the essential contribution of quadratic terms ($x_i x_j$) in equation 1. Because regression algorithms were developed for continuous variables,[25] a new specialized algorithm or theory for binary arrays may be needed in future work.

For regression, the database was split randomly into the training and testing datasets. The training dataset was used for the regression, and the testing dataset was used only for prediction. The $R^2$ score for the testing dataset was as high as 0.5 with a dimension of $x$ = 576 (Figure 3b; results with other conditions are shown in Figure S3). The regression scores were even higher than those obtained with random forest regression models, which are regularly used to treat nonlinear systems (Figure S4).[1, 5] Therefore, the proposed linear model with quadratic terms was sufficient to predict chemical properties from binary arrays.

The dimension of arrays, $n$, should be sufficiently larger than 100 to achieve high prediction accuracy (Figures S3). However, the corresponding search space of $>2^{100} \cong 10^{30}$ was too large for conventional computers. Thus, new machines are needed to explore best input $x_{ideal}$ effectively.

It was essential to obtain $x_{ideal}$ yielding the highest predicted performance, $y_{pred}$, from equation 1. A digital annealer was used to solve the problem (Figure 4a). As a control, simulated annealing on a conventional computer was examined. The annealing time with the annealer was only around 2.5 s, regardless of dimension $n$. In contrast, the required time with the conventional machine increased quasi-exponentially and exceeded 300 s with $n$ = 1250.

Moreover, $y_{pred}$ obtained in the control experiment was substantially smaller than in the annealer with $n$ > 100, demonstrating that the control could not reach the global maximum of $y_{pred}$ (Figure 4a). An impractically long time would be required to match the results of the annealer.

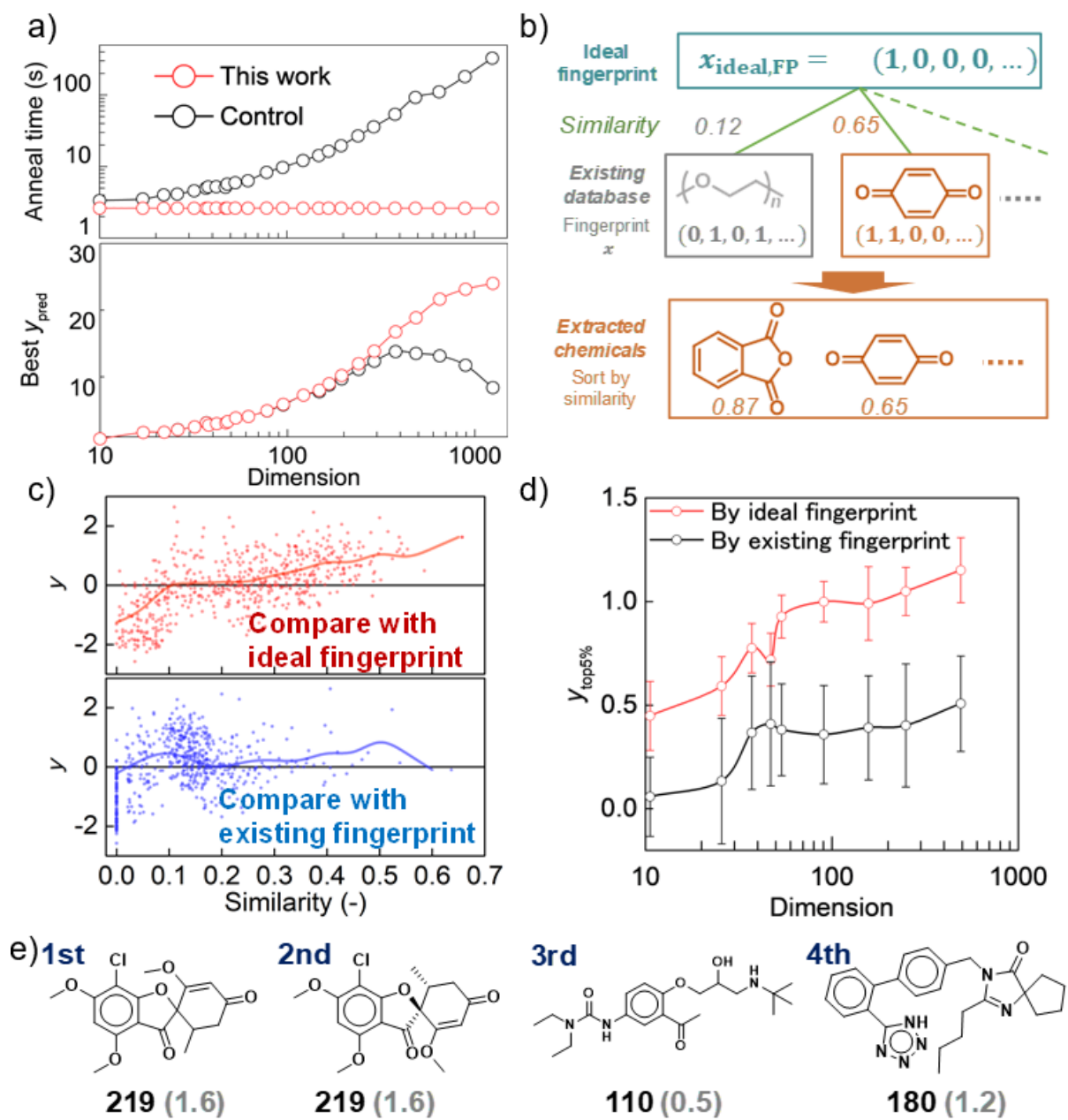

**Figure 4.** a) Annealing time and best $y$ found as a function of the dimension of $x$. Linear regression models were prepared with different dimensions. $x_{ideal}$ giving the highest $y_{pred}$ was explored by the digital annealer and a conventional computer as a control. b) Scheme to extract candidate compounds using similarity scores with $x_{ideal}$. c) Relationship between $y$ and similarity to a target fingerprint. In the upper panel, 20% of the test data was compared with $x_{ideal}$. In the bottom panel, the test dataset was compared with the existing fingerprint, which gave the highest *y* in the training dataset. The analysis was conducted with a dimension of 248. d) Statistical trends for the extraction performance. Averages *y* of the top 5% similar compounds are plotted as a function of dimension. Error bars show the standard deviation of the five-fold cross-validation results (see Experimental section for details).

After decompressing $x_{ideal}$ to the fingerprint form, $x_{ideal,FP}$, Tanimoto similarity scores[26] were calculated to compare the fingerprints of the ideal compound and those recorded in the test dataset (Figure 4b). Experimental *y* became statistically higher when the high similarity scores increased (Figure 4c, red plots). The increase indicated that $x_{ideal}$ found by the annealer maintained essential and universal information to achieve desirable performance. In contrast, there was no significant relationship between *y* and the similarity scores when the fingerprints of an existing compound with the highest *y* in the training dataset were compared (Figure 4c, blue plots). This was because the structural information of only one compound was too specific and narrow, and may have overlooked some essential structural information to achieve the desired performance. We calculated the average *y* of the top 5% most similar compounds (Figure 4d). Statistically, the annealing approach was preferable with various dimensions *n* rather than with existing fingerprints. Most of the extracted chemicals displayed *z*-scores of >1, corresponding to melting points of around 200 °C.

The proposed framework was useful for extracting chemicals with other desirable parameters, such as low melting temperature, high solubility parameter, high glass transition temperature, and high photoconversion efficiency in organic thin-film solar cells (Figures S5−S10). This versatile framework could be used for a wide range of materials and properties, not only for organic molecules, but also for inorganic compounds. During the annealing processes, additional restriction conditions using the fingerprints of existing chemicals also improved the extraction performance. The restrictions increased the structural feasibility of $x_{ideal}$, whereas only $y$ was pursued without the restriction (Figure S5, see Supporting Discussion for further information). Additional studies are needed to provide systematic insights into the extraction processes.

## 4. Generating new chemicals from ideal fingerprints

Instead of comparing the similarity scores with the existing database, directly generating molecules from $x_{ideal}$ is preferable for accessing the vast chemical search space. With a large dimension $n$ (e.g., >1000), $x$ can express a sufficiently large number of chemicals, $2^{1000} \cong 10^{300}$. Therefore, if appropriate mapping functions ($f_{MAP}$ and $f_{MAP}^{-1}$) are given, the annealing framework may be able to explore arbitrary organic molecules. However, determining the functions, especially the inverse functions, is still unwieldy, regardless of the remarkable progress in chemoinformatics and machine learning.[1, 11, 13]

Here, we used a more straightforward approach to preparing chemicals directly from $x_{ideal}$. After decompressing $x_{ideal}$ to the 2048-dimensional fingerprint, $x_{ideal,FP}$, chemical structures were generated from the fingerprint fragments (Figure 5a). A fingerprint bit $x_{FP,i}$ represents a specific fragment of a chemical structure. In the example in Figure 5a, bit $x_{FP,0}$ indicates whether an aromatic ring is available in a molecule. The bit is 1 if the ring is present and else 0.

From the melting temperature database, chemical fragments were extracted by the breaking of retrosynthetically interesting chemical substructures (BRICS) algorithm.[27] Then, some fragments sharing the same bit of the ideal fragment were extracted and reconnected to make molecules according to BRICS (Figures 5b, 5c, and S11).

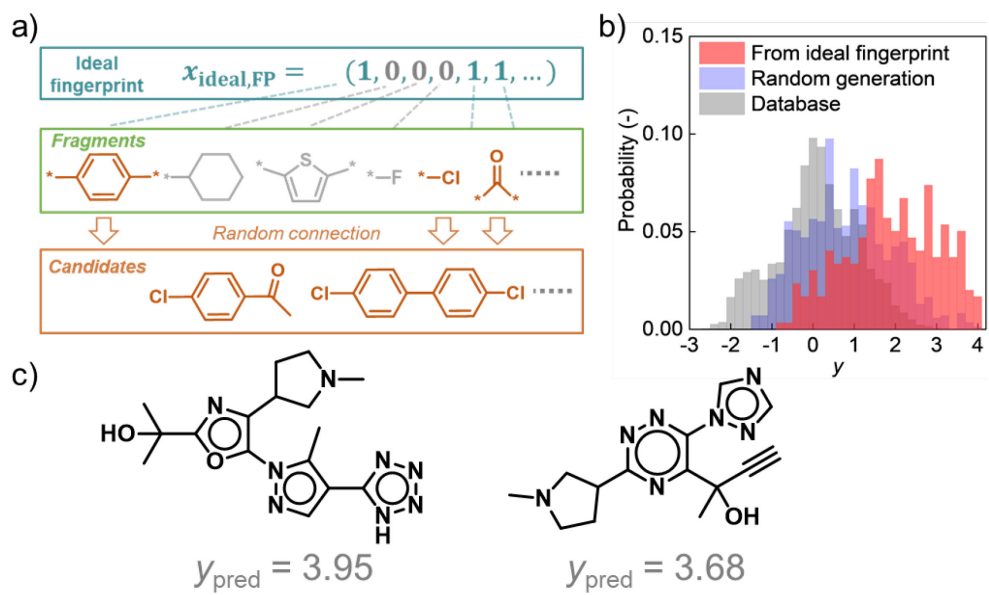

**Figure 5.** a) Framework for generating new chemicals from the ideal fingerprint, $x_{\text{ideal}}$. b) Probability distribution of the standardized melting points from the original database, and for molecules generated from the ideal fingerprint and via a random process. Predicted values ($y_{\text{pred}}$) are shown with the generated molecules. c) Examples of the generated chemicals.

Predicted $y$ of the generated chemicals was statistically higher than the chemicals of the original database and randomly generated chemicals (control, Figure 5b). The average scores were 1.80, 0, and 0.71 for the proposed approach, original database, and random generation, respectively. Large values of ca. 4 were even obtained from the ideal fingerprint (Figures 5c and S11). Our high-throughput molecular generation system will help researchers design new materials with desirable properties.

## 5. Optimizing the composition of Li$^+$-conducting electrolytes

Finally, we extended the annealing framework to find the optimal compositions of multiple chemicals forming composite materials. The search space increases exponentially with *N*-component systems (*N* > 1). We focused on finding the best composition of organic Li$^+$-conducting electrolytes consisting of up to six chemicals (Figures 6a and S12).

Li$^+$-conducting electrolytes were selected because of the essential role of Li$^+$ batteries in society and the demand for higher conductivity to improve outputs.[28] Most electrolytes consist of at least two components, namely, a solvent and salt.[3, 29-31] Additional chemicals are often added to improve viscosity, conductivity, stability, and other battery properties.[29, 31] The search space becomes even larger when polymeric electrolytes are considered as essential components for next-generation solid-state batteries and conventional gel electrolytes.[3, 30]

We have previously explored organic Li$^+$-conducting electrolytes using machine learning.[3] The largest database of Li$^+$-conducting electrolytes was prepared, which covered both monomeric and polymeric conductors (around unique 1200 records near room temperature, Table S1 and Figure S13). The prediction accuracy of a model trained with the database was as accurate as that of experienced researchers[3] and the machine learning model identified new polymeric conductors. However, a narrow search space (<10,000) was used during the screening and it is likely that candidates in the broader search space were overlooked.

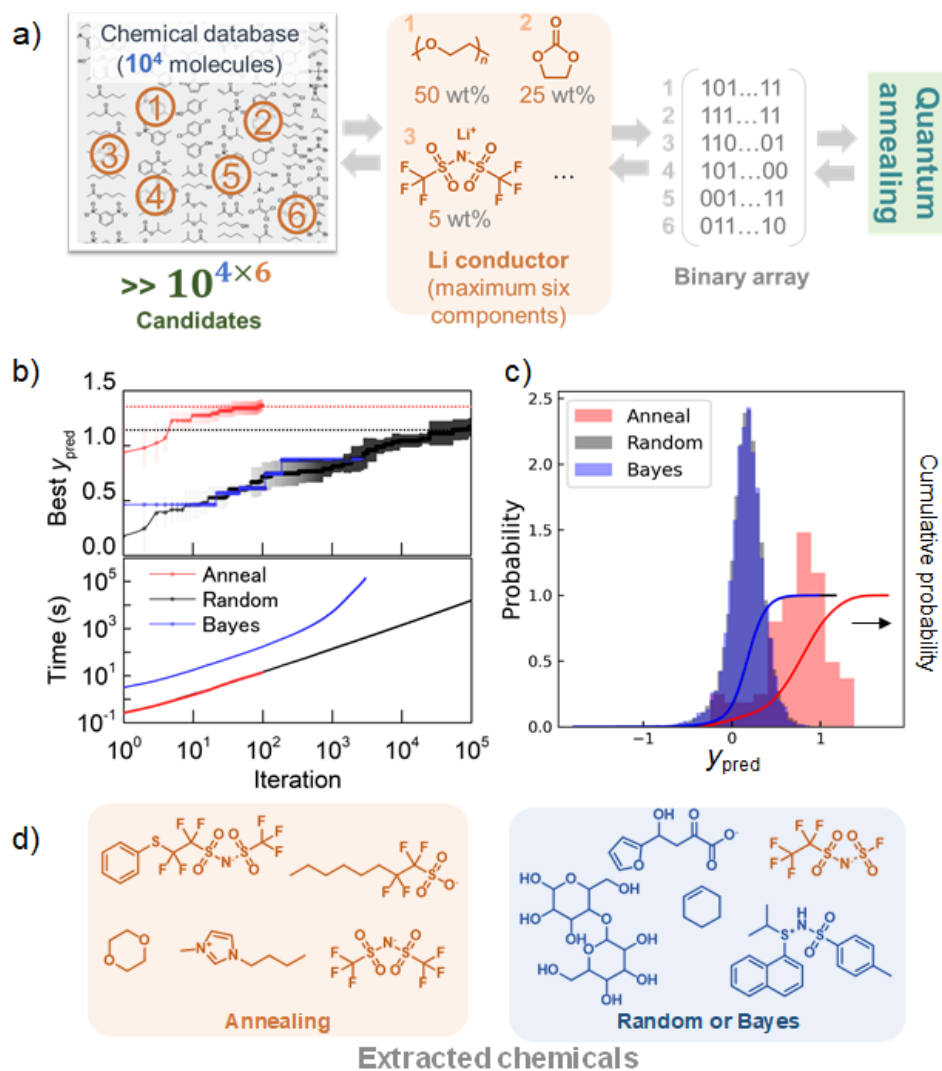

**Figure 6.** a) Scheme to treat composite materials. b) Results for exploring optimal composites by the annealing, random sampling, and Bayes optimization approaches. The average scores for five independent trials are shown in the cases of the anneal and random approaches. Error bars represent standard deviations. Only one trial was conducted for the Bayes, because of the excessively long calculation time. The dimension of *x* was 398. c) Probability distribution of the composites extracted by the three approaches. d) Example structures of the extracted chemicals.

Composition information about electrolytes was converted into binary arrays. For simplicity, up to six chemical components[3] were considered for one composite. Weight ratios of the components were also included in $x$. The continuous values were converted into binaries by a unary method (see Experimental section for details). Then, regression was conducted to predicted room temperature conductivity. A linear model yielded high $R^2$ scores of 0.88 and 0.68 for the training and testing datasets, respectively (Figure S13b).

After obtaining $x_{ideal}$ with the annealer from the regression model, the decompressed array was compared with the fingerprints of a chemical compound database. There were around 10,000 candidate chemicals in the database. Therefore, the search space was as big as $10^{4 \times 6}$, ignoring weight ratios (i.e., an additional five independent variables).

Compounds were extracted from the database by comparing their fingerprints with those of the ideal compounds. The selections were made by a random process, where high-similarity compounds were extracted preferentially according to a Gaussian distribution. The weight ratio was calculated by restoring the corresponding binaries from $x_{ideal}$. The selection was repeated to prepare 100 types of candidate composite (Figure 6b). These selection processes were executed five times to ensure statistical validity. On average, it took only around 3 s to obtain composites with $y_{pred} > 1.3$. After finishing the 100 iterations, the maximum value increased to 1.4, which took about 14 s.

However, $10^5$ iterations were not sufficient to achieve $y_{pred} = 1.3$ when the compounds and weight ratios were selected by a completely random process (Figure 6b). The average time to reach the maximum ($y_{pred} = 1.17$) was 15,600 s, whereas the annealing method took less than 0.85 s. Therefore, the latter was at least $15,600/0.85 \cong 18,000$ times faster than the standard approach.

Bayesian optimization, a conventional approach to explore optimal conditions,[1, 32] did not find the best combination (Figure 6b). The best score was only 0.87 after 4000 iterations, although the method took an exceptionally long time of 356,000 s. The poor result was attributed to the non-smoothness of the function and search space that was too large.

Smoothness is essential to finding the global maximum (or minimum) effectively by Bayesian optimization.[32] However, $y_{pred}$ changed drastically only when a single component of a composite was changed (Figure S14) because the compound ID in the database was a categorical variable, not a molecular descriptor.[1] At this point, the extraction performance of the Bayesian process was similar to that of the random process (see histogram of $y_{pred}$ in Figure 6c). It took an excessive amount of time to fit the responses to the fluctuating responses with a costly stochastic model during Bayesian optimization.[32]

The annealing approach improved the statistical distribution of $y_{pred}$ greatly. For instance, 7% of the proposed electrolytes provided $y_{pred}$ of over 1.2, whereas in the random approach only 0.0002% did (Figure 6c). The speedup ratio for this specific task was as large as $4 \times 10^4$. The proposed system finished the task in only 14 s, whereas the control method would have needed $5.5 \times 10^5 \text{s} \cong 6$ days. The Bayes method could not yield an electrolyte of $y_{pred} > 1.2$. A simple comparison of the iteration rates (7.2 and 0.011 compounds per seconds for the annealing and Bayes, respectively) indicated that the annealing would be at least $4 \times 10^4 \times (7.2/0.011) = 3 \times 10^7$ times faster than the Bayes approach.

From the viewpoint of actual experiments, a sufficiently large number of candidates are needed. The materials prescreened by computers are still candidates extracted from $f_{ML}$. Because the predictions are not perfect, most candidates would be excluded during the next screening steps (e.g., human-based composition check and real experiments). The poor extraction efficiency and excessively long calculation time of the conventional methods may not be useful for practical projects.

The quality of the extracted compounds obtained by annealing was also higher than that of the controls (Figures 6d, S15−S17). About half of the chemicals extracted by annealing were experimentally examined as electrolyte components for Li$^+$ batteries[3, 29] (shown in yellow, Figure S15). In contrast, 75% or more of the compounds extracted by the controls were unconventional (Figures S16 and S17). Experimentally, most of the compounds proposed by the random approach would not provide ionic conductivity because the components were far from salt- or solvent-like structures (e.g., no polar solvents in a composite, Figures S16 and S17).

High prediction scores could be obtained by $f_{ML}$ even with noisy composites, whose structures were dissimilar from electrolytes, because of the lack of training data during machine learning. We trained the model only with Li$^+$-conducting electrolytes, not with non-Li$^+$-conducting composites. This was because current machine learning models can process only specific tasks in narrow fields due to the limited complexity of the model and computing power,[1] and adding too broad a context may decrease the prediction accuracy.[33] The preparation of databases, which is generally a manual process,[3, 5] is also hugely time-consuming. Therefore, the inaccurate predictions by $f_{ML}$ are currently unavoidable and should be avoided during the candidate selection step.

The ratio of conventional and unconventional compounds in proposals is crucial during screening. If there are too many unconventional compounds, the proposed compounds will be "noisy" (Figures S16 and S17), whereas suggesting too many conventional compounds hinder the discovery of new materials. The ratio of 50% provided by annealing may be acceptable during the exploration of potential candidates. The balance can be tuned by adding a restriction condition during annealing in a similar way to the single-component system (Figure S5).

## 6. Future challenges

Although the proposed annealing system is superior to conventional approaches, there are still problems that must be addressed. The prediction accuracy of $f_{ML}$ is not sufficient. Based on standard knowledge of electrochemistry, we suspect that most of the proposed unconventional compounds, even those obtained by the annealing approach, will not improve sufficient conductivity (Figure S15).[31] Apart from experiments or costly simulations, there is no way to confirm whether the proposed unconventional compounds are unexpectedly good or are noise.

Regardless of the higher accuracy required, $f_{ML}$ is currently limited to the linear Ising model (equation 1). Models that are more sophisticated may be needed to improve accuracy. Hybrid calculations with conventional computers are needed to implement more complex functions while awaiting the development of gate-based quantum computers.[18, 19] In addition, the mapping functions to convert material information reversibly to binary arrays ($f_{MAP}$ and $f_{MAP}^{-1}$) must be optimized to access the enormous search space more efficiently.

## 7. Conclusion

We constructed a rapid extraction system for chemicals using machine learning and a quantum-inspired annealing machine. The framework allowed rapid exploration of chemicals and compositions, and the extraction was at least $10^4$–$10^7$ times faster than conventional methods. Increasing the speed of prediction methods is essential to tackle the combinatorial explosion problem in chemical screening (e.g., $10^{60N}$ candidates for an $N$-component system of small organic molecules). The structures extracted by our method were higher quality than those extracted by the control methods because the annealer could extract the essential features of desirable compounds. The screening system will be critical to dramatically shortening the time required to optimize versatile, functional materials and devices.

**Acknowledgments**

This work was partially supported by Grants-in-Aid for Scientific Research (Nos. 17H03072, 18K19120, 18H05515, 20H05298, 19K15638, and 20H05298) from MEXT, Japan. The work was also partially supported by the Research Institute for Science and Engineering, Waseda University.

**Competing interests**

The authors declare no competing interests.

Supporting Information

**Tackling the challenge of a huge materials science search space with quantum-inspired annealing**

*Kan Hatakeyama-Sato, Takahiro Kashikawa, Koichi Kimura, and Kenichi Oyaizu\**

**Methods**

**1. General information**

Program code (written in Python 3) and databases used in this paper can be accessed at https://github.com/KanHatakeyama/annealing_project. Annealing was conducted using digital annealing unit 2 (DAU).[1] Another calculation was done using an Intel Xeon Gold 6148 CPU @ 2.40 GHz.

**2. Databases**

Five types of databases were introduced in this study (Table S1).

a) Bradley's dataset[2]

The database contains the molecular structures of about 3000 types of organic chemicals and their melting points. Molecular structures and melting points were set as *x* and *y*, respectively.

b) Delaney' dataset[3]

The database contains properties of about 1100 organic chemicals. Molecular structures and their experimental solubility parameters were set as *x* and *y*, respectively.

c) Polymer database[4]

The database contains glass transition temperatures of about 170 conventional polymers. Repeating polymer unit structures and the transition temperatures were set as *x* and *y*, respectively.

d) Organic solar cells[5]

The database records the performances of organic thin-film solar cells and contains the experimental responses of about 1200 cells. Molecular structures of the donor molecules and the photoconversion efficiency of the corresponding cell were set as *x* and *y*, respectively. Other parameters that may contribute to the efficiency, such as molecular weight, were ignored for simplicity.

e) $Li^+$-conducting electrolytes[6]

This database contains various types of organic liquid- and solid-state $Li^+$-conducting electrolytes. Data for around 1200 unique electrolytes are recorded around room temperature. Typically an electrolyte consists of multiple chemicals. Chemical structures (up to six compounds) and their weight ratio were set as *x*, and room temperature ionic conductivity was used as *y*. Other parameters, including inorganic additives and polymer structures (e.g., cross-linking), were excluded from *x* for simplicity.

**3. Scheme to explore chemicals for one-component system**

Except for $Li^+$-conducting electrolytes, chemical structures were explored using DAU according to the following scheme (Figure S1).

1) Data splitting

The original database was split into training and testing datasets randomly. In Figures 4d, S3, S4, and S6−S10, five-fold cross-validation was used to ensure statistical validity.

2) Calculation of 2048-dimensional chemical fingerprint

All molecular structures were recorded by character strings based on a simplified molecular-input line-entry system (SMILES). The chemical information was converted into 2048-dimensional extended connectivity fingerprint, $x_{FP}$, using a free library, RDKit (equation S1).[7,8]

$$x_{FP} = (x_1, x_2, x_3, \ldots, x_{2048}) \tag{S1}$$

3) Fingerprint compression

The generated fingerprints were compressed into $n$-dimensional binary arrays, $x$ (Figure S2 and equation S2).

$$x = (x_1, x_2, x_3, \ldots, x_n) \tag{S2}$$

The variations of $x_{FP,i}$ ($i$ = 1, 2, …, 2048) in each record were calculated for the training dataset. If $x_{FP,i}$ had a slight variation (e.g., 0,0,1,0,…0,0,0 or 1,1,1,1….1,0,1 for each record), it was excluded from $x$. The threshold was changed manually to generate $x$ with various dimensions. During decompression, the removed $x_{FP,i}$ value was restored using the mode value.

4) Addition of interactions

Interactions were added to obtain binary arrays, $x_{int}$, for machine learning (equation S3).

$$\begin{aligned}x_{int} = (&x_1, x_2, x_3, \ldots, x_n, \\ &x_1 x_2, x_1 x_3, \ldots, x_1 x_n, \\ &x_2 x_3, x_2 x_4, \ldots, x_2 x_n, \\ &, \ldots, x_{n-1} x_n)\end{aligned} \tag{S3}$$

5) Training model

A linear regression model $f_{ML}$ was built to predict $y$ from $x$ for the training dataset (equation S4). Target parameters were converted into $z$-scores to obtain $y$.

$$y_{\text{pred}} = f_{\text{ML}}(x) = \sum_{i \neq j} J_{ij}\, x_i x_j + \sum_{i} h_i\, x_i \qquad (S4)$$

The coefficients $J_{ij}$ and $h_i$ were determined by training. The models were prepared by stochastic gradient descent (SGD) learning, using a module of the scikit-learn library.[9] SGD learning was used because the quadratic terms ($x_i x_j$) increased the dimension of the inputs, and thus increased calculation time enormously. The Ridge regression with a regularization strength $\alpha = 0.1$ was used unless noted otherwise. For the control experiments, random forest repressors (scikit-learn) were trained with $x$ without adding quadratic terms.

6) Calculation of $R^2$ score

$R^2$ scores were calculated for the actual ($y$) and predicted values ($y_{\text{pred}}$) of the records in the training and testing datasets. The testing datasets were used only for prediction, not for training the model.

7) Finding ideal binary $x_{\text{ideal}}$ using DAU

The model was equivalent to the quantum Ising model, and thus could be processed by annealers to find the global minimum. The task is called the quadratic unconstrained binary optimization problem (QUBO).[1] The QUBO matrix, $M_{\text{QUBO}}$, was passed to DAU to find $x_{\text{ideal}}$, giving $y_{\text{pred}} \cong \text{argmin}\,(f_{\text{ML}}(x))$ (equation S5). The maximum was explored by passing $-M_{\text{QUBO}}$ to the machine. For the control experiments, regular computer simulated annealing was conducted using an open-source module (Blueqat, Figure 4a).[10]

$$M_{\text{QUBO}} = \begin{pmatrix} h_1 & J_{12} & J_{13} & J_{14} & \cdots \\ 0 & h_2 & J_{23} & J_{24} & \cdots \\ 0 & 0 & h_3 & J_{34} & \cdots \\ 0 & 0 & 0 & h_4 & \cdots \\ \cdots & \cdots & \cdots & \cdots & \cdots \end{pmatrix} \qquad (S5)$$

In Figures S6–S10, the QUBO matrix was biased with a diagonal matrix with a perturbation coefficient of $c_{\text{pert}}$ (equation S6).

$$= M_{\text{QUBO}} + c_{\text{pert}} \begin{pmatrix} b_1 & 0 & 0 & 0 & \cdots \\ 0 & b_2 & 0 & 0 & \cdots \\ 0 & 0 & b_3 & 0 & \cdots \\ 0 & 0 & 0 & b_4 & \cdots \\ \cdots & \cdots & \cdots & \cdots & \cdots \end{pmatrix} \tag{S6}$$

where

$$b_i = \begin{cases} 1 & \text{if } x_{\text{exist},i} = 0 \\ -1 & \text{if } x_{\text{exist},i} = 1 \end{cases}$$

Here, $x_{\text{exist}}$ was $x$ that gave the highest $y$ in the training dataset.

8) Calculation of similarity

The $n$-dimensional array, $x_{\text{ideal}}$, was restored to a 2048-dimensional fingerprint, $x_{\text{ideal,FP}}$. Then, Tanimoto similarity scores were calculated for each compound in the test dataset against $x_{\text{ideal,FP}}$.

9) Extraction of chemicals with high-similarity compounds

Compounds in the test dataset were sorted according to their similarity scores. The average $y$ of the compounds with the highest 5% similarity was defined as $y_{\text{top5\%}}$. For the control, results are also shown using a restored fingerprint, $x_{\text{exist,FP}}$, instead of $x_{\text{ideal,FP}}$ in Figure 4c.

10) Generation of new chemicals

New chemicals were generated directly from $x_{\text{ideal,FP}}$ (Figure 5). RDKit modules were used to make molecules. Chemicals in the corresponding database were fragmented by the breaking of retrosynthetically interesting chemical substructures (BRICS) algorithm. Fragments with a specific fingerprint value of $x_{\text{FP},j}$ (= 1) were extracted. Here, $j$ could be the integers satisfying $x_{\text{ideal,FP},j} = 1$. The extracted fragments were reconnected randomly according to the BRICS algorithm. As a control, all types of fragments were reconnected randomly (random generation, Figure 5b).

**4. Scheme to explore chemical composites**

Chemical components of Li$^+$-conducting electrolytes were explored using DAU according to the following scheme (Figure S12). Most procedures before converting into binaries were the same as in our previous report.[6]

1) Load database

The database consisted of compound and composite databases. The compound database maintained the chemical structures of organic molecules, including polymers and normal low-molecular-weight molecules. Li atoms were omitted because the database focused on only Li$^+$ electrolytes. For convenience, bonds representing repeating or grafting units were expressed by Mg and Ca atoms, respectively. Conductivity data around room temperature were extracted. Conductivity was converted in a logarithmic scale, and then into a $z$-score.

2) Calculation of the weight ratio of the components

In the databases, chemical compositions were recorded as either molar or weight ratio. All compositions were recalculated as weight ratios.

3) Compound sorting by weight ratio

For the unique expression of the database, the recorded compounds were sorted by their weight ratio (from large to small). If there were more than six components, only the top six weight ratio components were used for data processing for simplicity.

4) Calculation of fingerprints

For one composite, 2048-dimensional fingerprints of the components were calculated.

5) Conversion of weight ratio into a binary array

The composition ratios of chemicals, which were continuous values, were expressed by binary arrays. First, the original weight ratio $r$ was normalized to $s$.

$$r = (r_1, r_2, r_3, \ldots, r_6) \tag{S7}$$

$$\begin{aligned}s &= (s_1, s_2, s_3, \ldots, s_5) \\ &= (\log(\frac{r_1}{r_2}), \log\left(\frac{r_2}{r_3}\right), \ldots, \log(\frac{x_5}{x_6}))\end{aligned} \tag{S8}$$

The normalized form, $r_i/r_{i+1}$ ($\geq 1$), provided a restriction condition during the composition screening. The restriction of $s_i \geq 1$ was easier to implement than the original condition ($r_1 + r_2 + \cdots + r_6 = 1$ and $r_1 > r_2 > \cdots > r_6 > 0$). The values were converted on a logarithmic scale because of their large scale ranges.

The normalized values, $s_i$, were approximated with binaries, $b_j = 0$ or 1 (unary method).

$$s_i = v_{\min} + \frac{v_{\max} - v_{\min}}{N_{\text{bin}}} \sum_j^{N_{\text{bin}}} b_j \tag{S9}$$

Here, $v_{\max}$ and $v_{\min}$ were the maximum and minimum values of $s_i$ observed in the database. The number of binaries used, $N_{\text{bin}}$, was set as 10. The binary method ($\sum_j^N 2^{j-1} b_j$) was not compatible with the following linear regression and annealing because its dynamic range was too large.

6) Split dataset annealing

The binary database was split into the training and testing datasets (9/1 ratio) and compressed to $n$-dimensional arrays ($= x$). The dimension was always set to be 398 for the electrolyte system. The fingerprint and weight ratio information was included in $x$. Polymer structures, molecular weight, inorganic additives, and other related parameters[6] were excluded for simplicity. Up to annealing, the same procedures were applied in the same way as in the single-component system.

7) Calculation of similarity to decompressed $x_{ideal}$

A maximum of six types of ideal fingerprints could be obtained from decompressed $x_{ideal}$. The components were explored by referencing a chemical compound database, containing about 10,000 SMILES structures from various sources. Chemicals with higher similarities were selected as the components. The selection was made randomly using sampling with a Gaussian distribution. The weight ratio was restored from equation S9. The random component selection was repeated 100 times (Figure 6b). For the control experiments, chemicals and weight ratios were selected completely randomly or by Bayesian optimization using a Python library of GPyOpt.[11]

**Table S1.** Information about databases used for analyses.

| Name | Input ($x$) | Target property ($y$) | Number of records | Ref. |
|---|---|---|---|---|
| Bradley's dataset | Molecular structure | Melting point | 3000 | [2] |
| Delaney' dataset | Molecular structure | Solubility parameter | 1100 | [3] |
| Polymer database | Molecular structure | Glass transition temperature | 170 | [4] |
| Organic solar cells | Molecular structure | Photoconversion efficiency | 1200 | [5] |
| Li$^+$-conducting electrolytes | Multiple molecules (including polymers) | Ionic conductivity | 1200[a] | [6] |

[a]Number of records around room temperature.

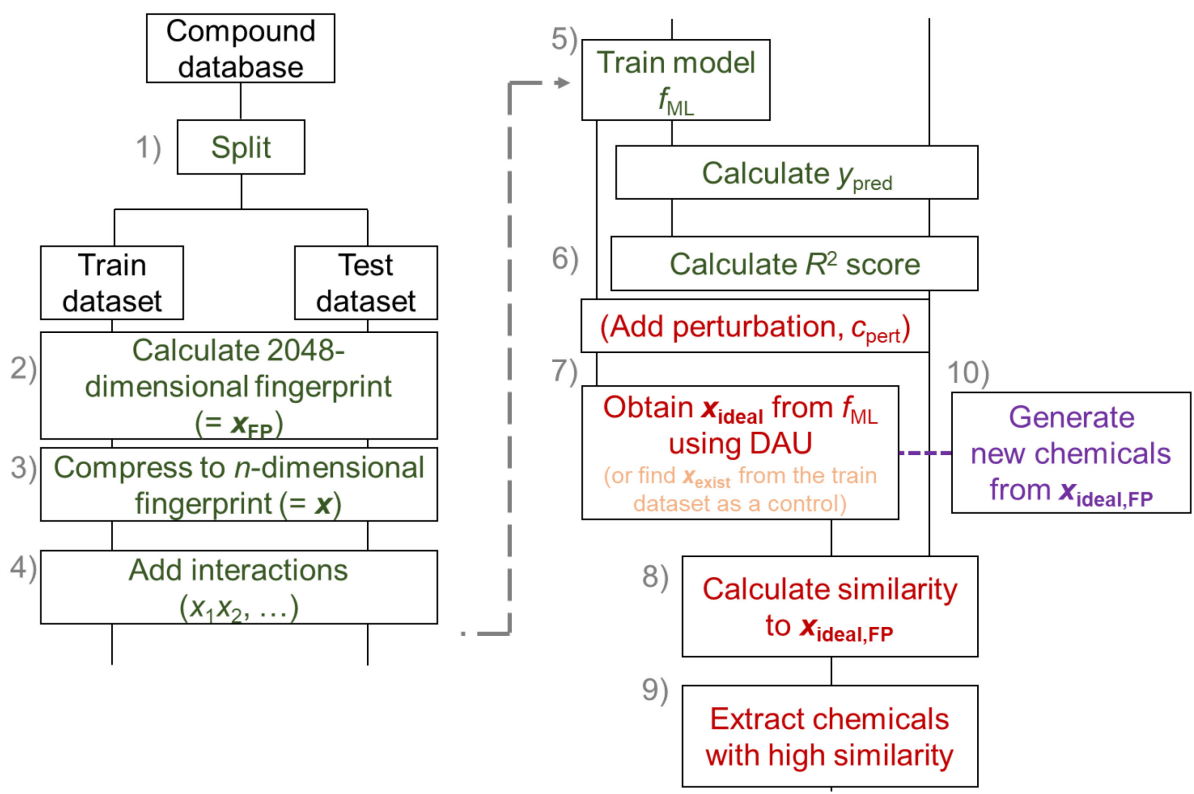

**Figure S1.** General scheme for exploring chemicals using DAU (one-component system).

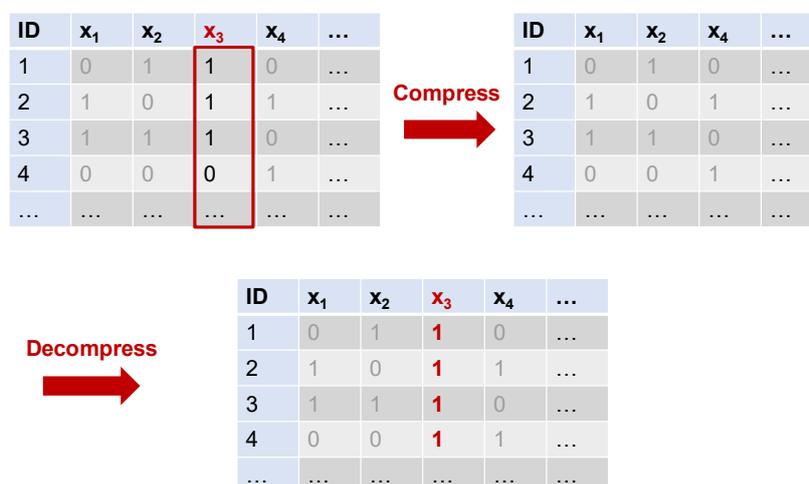

**Figure S2.** Compression and decompression of the binary arrays. In the figure, $x_3$ was removed during compression because it gave 1 in most cases. During decompression, $x_3$ was restored using the mode value of 1. Therefore, the compression/decompression processes are irreversible. The compression was done with any $x_i$ with a variance was less than the threshold.

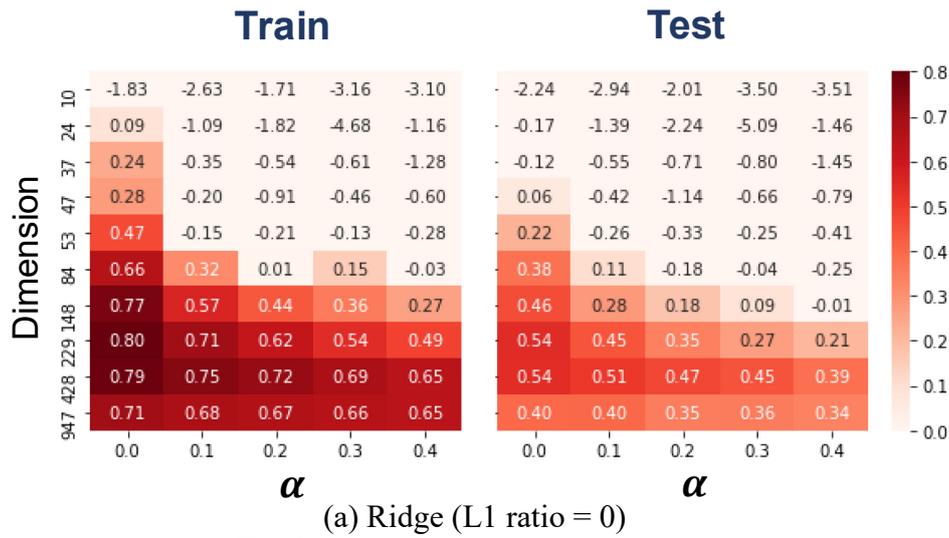

(a) Ridge (L1 ratio = 0)

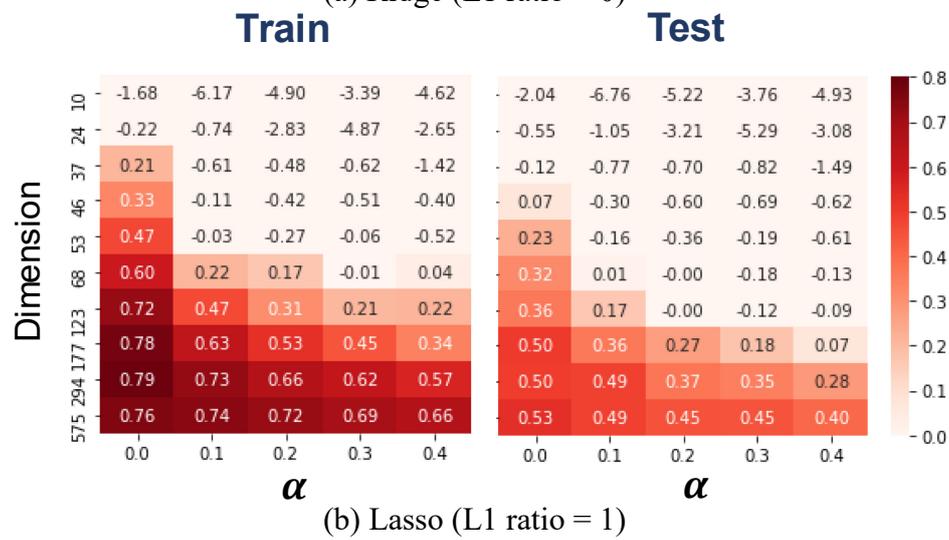

(b) Lasso (L1 ratio = 1)

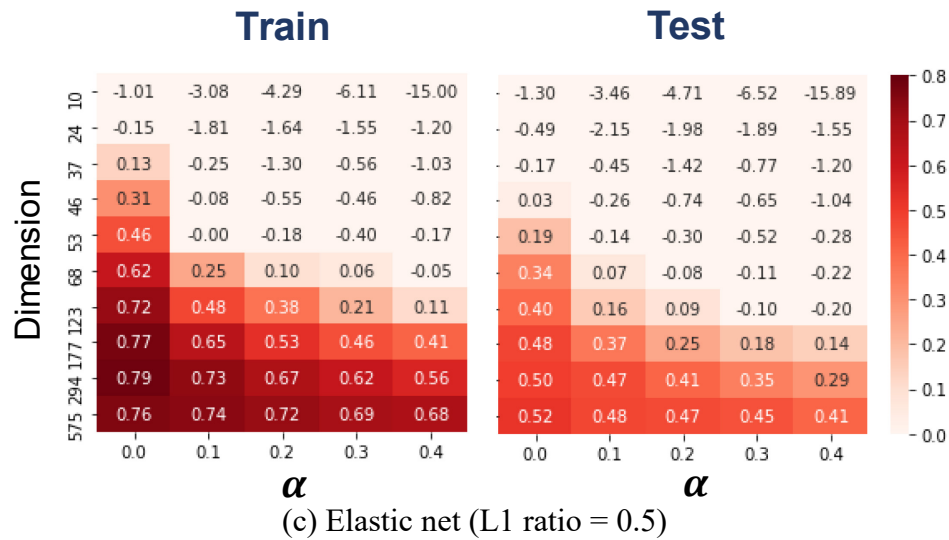

(c) Elastic net (L1 ratio = 0.5)

**Figure S3.** Grid search results for optimizing the hyperparameters for linear regressions. a) Ridge model, b) Lasso model, and c) elastic net model. Five-fold cross-validation was conducted. Average $R^2$ values for the training and testing datasets are shown. Dimension, regularization strength $\alpha$, and L1 ratio were changed. Regression with $\alpha = 0$ gave the best $R^2$ score but should be avoided for machine learning because of the multicollinearity problem.

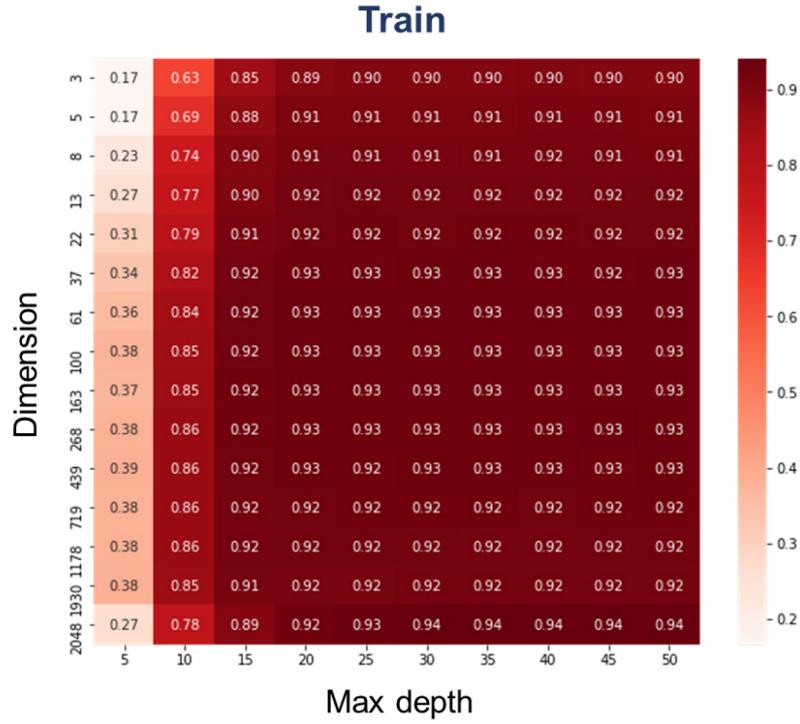

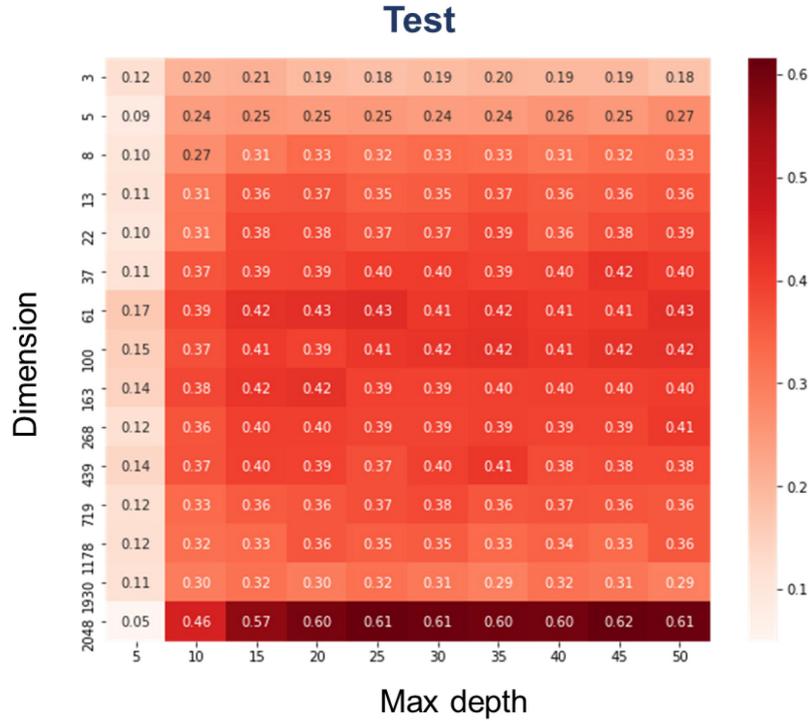

**Figure S 4.** Grid search results for optimizing the hyperparameters for random forest regressions. Five-fold cross-validation was conducted. Average $R^2$ values for the training and testing datasets are shown. The dimension and maximum depth of the trees were changed.

**Supporting discussion**

**Extraction of chemicals with versatile parameters**

In addition to compounds with a high melting temperature, compounds with a low melting temperature, high solubility parameter, high glass transition temperature, and high photoconversion efficiency were explored (Table S1, Figures S6−S10). Except for the last parameter, single-molecule properties were targeted. On the other hand, the photoconversion efficiency of organic thin-film solar cells can be determined by many various parameters, such as cell configuration.[5]

$R^2$ scores for the test datasets were as high as 0.5 with a sufficiently large $n > 100$ for most cases. Only the scores of solar cells were around 0.3, indicating that it was necessary to add other explanatory variables for regression.

A perturbation parameter, $c_{\text{pert}}$, was changed during the annealing process (Figure S5, equations S5 and S6). Standard annealing, discussed in the main manuscript, was done with $c_{\text{pert}} = 0$. As the coefficient increased, the binary obtained from the annealer would converge to $x_{\text{exist}}$, corresponding to the structure of an existing chemical, which gave the highest $y$ in the training dataset. The coefficient was given to increase the feasibility of the proposed chemical structures as a restriction condition.

In the high-melting-temperature exploration, the extraction score, $y_{\text{top5\%}}$ (average $y$ of the top 5% most similar compounds to the found binary), was high with smaller $c_{\text{pert}}$, indicating that this perturbation was unnecessary. On the other hand, the feasibility of $x_{\text{ideal}}$ may be questionable for $y_{\text{pred}}$. The observed $y_{\text{pred}}$ of >10 in Figure 4a was too large as a $z$-score. Although it did not affect the extraction efficiency of chemicals substantially, more feasible predictions may be required using restriction conditions (e.g., $c_{\text{pert}} > 0$).

When low-melting-temperature compounds were explored, the extraction was not successful with $c_{\text{pert}} = 0$ (Figure S7). The reason for this was unclear, but the lack of molecular feasibility may have affected the result. On the other hand, the extraction score $y_{\text{top5\%}}$ increased with $c_{pert} > 0$ and gave the maximum at $c_{pert} \cong 0.1$. The mixed properties of the ideal ($x_{ideal}$) and actual ($x_{exist}$) structure were needed for this extraction task.

For the glass transition temperature, mixed trends for high-melting-temperature compounds were detected (Figure S9). The small number of records in the database (= 170) made it difficult to extract reliable statistical trends, as the large error bars indicated.

For the compounds and solar cells with high solubility parameters, $y_{top5\%}$ reached the maximum at $c_{pert} \cong 0.1$ (Figures S8 and S10). The maximum was also observed at $c_{pert} = 1$ with some dimensions. Further studies are needed to understand the complex responses by systematically considering the effects of the fingerprint algorithms, regression model, similarity functions, functional groups, and database formats.

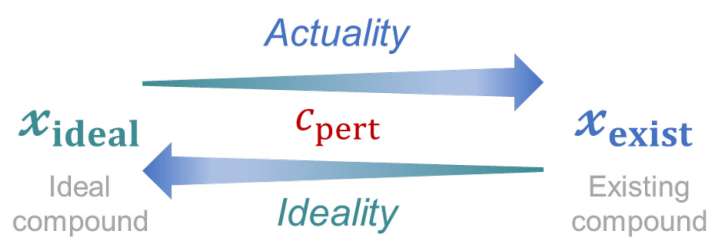

**Figure S5.** Role of $c_{pert}$ (≥0). The binary found by the annealer will converge to $x_{ideal}$ and $x_{exist}$ with smaller and larger $c_{pert}$, respectively.

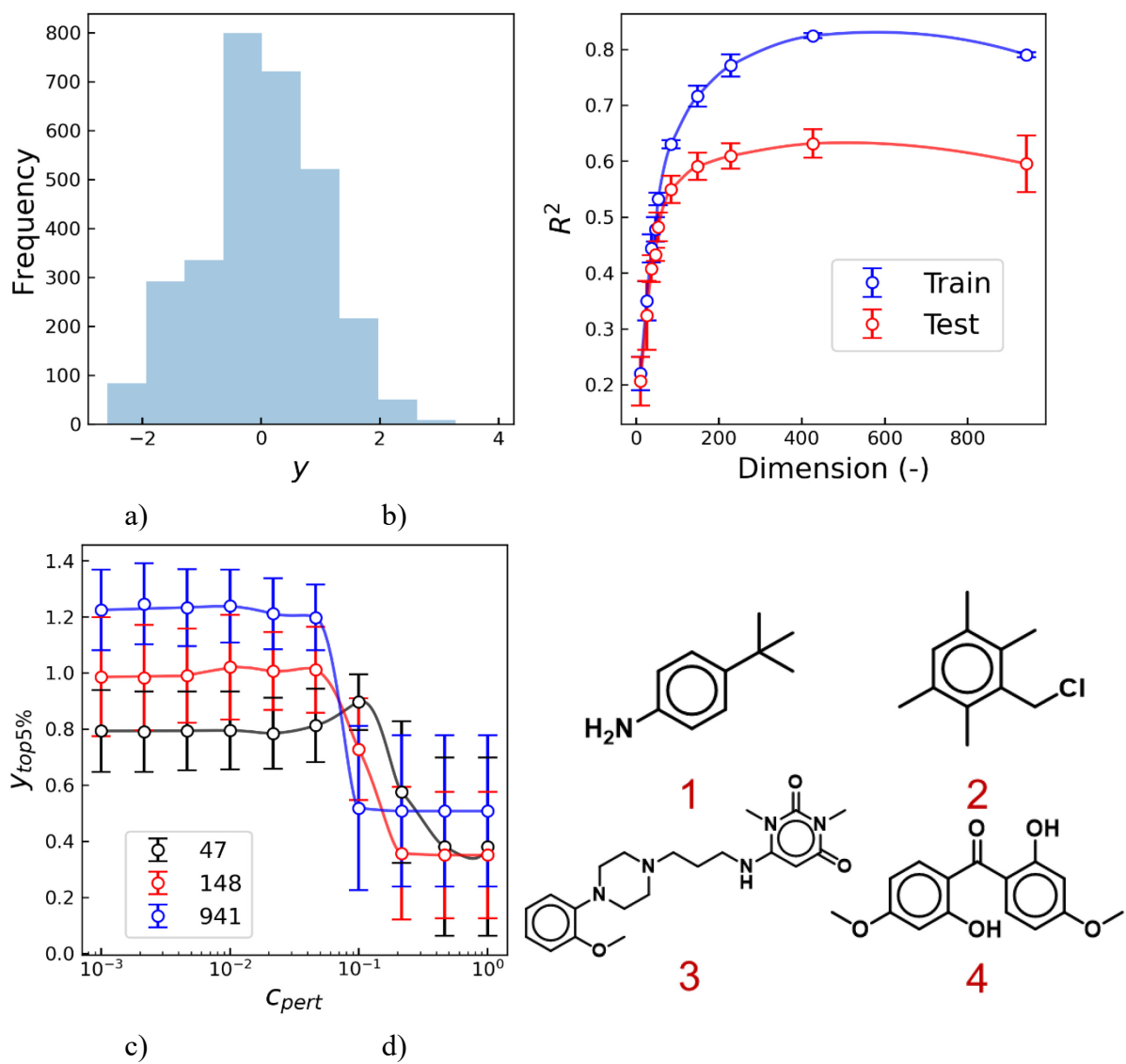

| TOP | SMILES | y | Similarity |
|---|---|---|---|
| 1 | CC(C)(C)c1ccc(N)cc1 | 1.6 | 0.57 |
| 2 | ClCc1c(c(cc(c1C)C)C)C | 1.4 | 0.49 |
| 3 | COc1ccccc1N3CCN(CCCNc2cc(=O)n(C)c(=O)n2C)CC3 | 1.4 | 0.48 |
| 4 | COc1ccc(c(c1)O)C(=O)c2ccc(cc2O)OC | 0.4 | 0.44 |
| 5 | CC(C)N(CCC(c1ccccn1)(c2ccccc2)C(N)=O)C(C)C | 0.4 | 0.44 |

e)

**Figure S6.** Exploration results for high-melting-temperature compounds by DAU.
a) Histogram of $y$ for the database. b) Regression results for the training and testing datasets. The dimension of $x$ was changed. c) Average $y$ of the top 5% of chemicals extracted by this approach. Dimension and $c_{pert}$ were changed. Values are the averages after five-fold cross-validation. Error bars represent standard deviation. d) and e) Information on the example structures of extracted chemicals with high similarities.

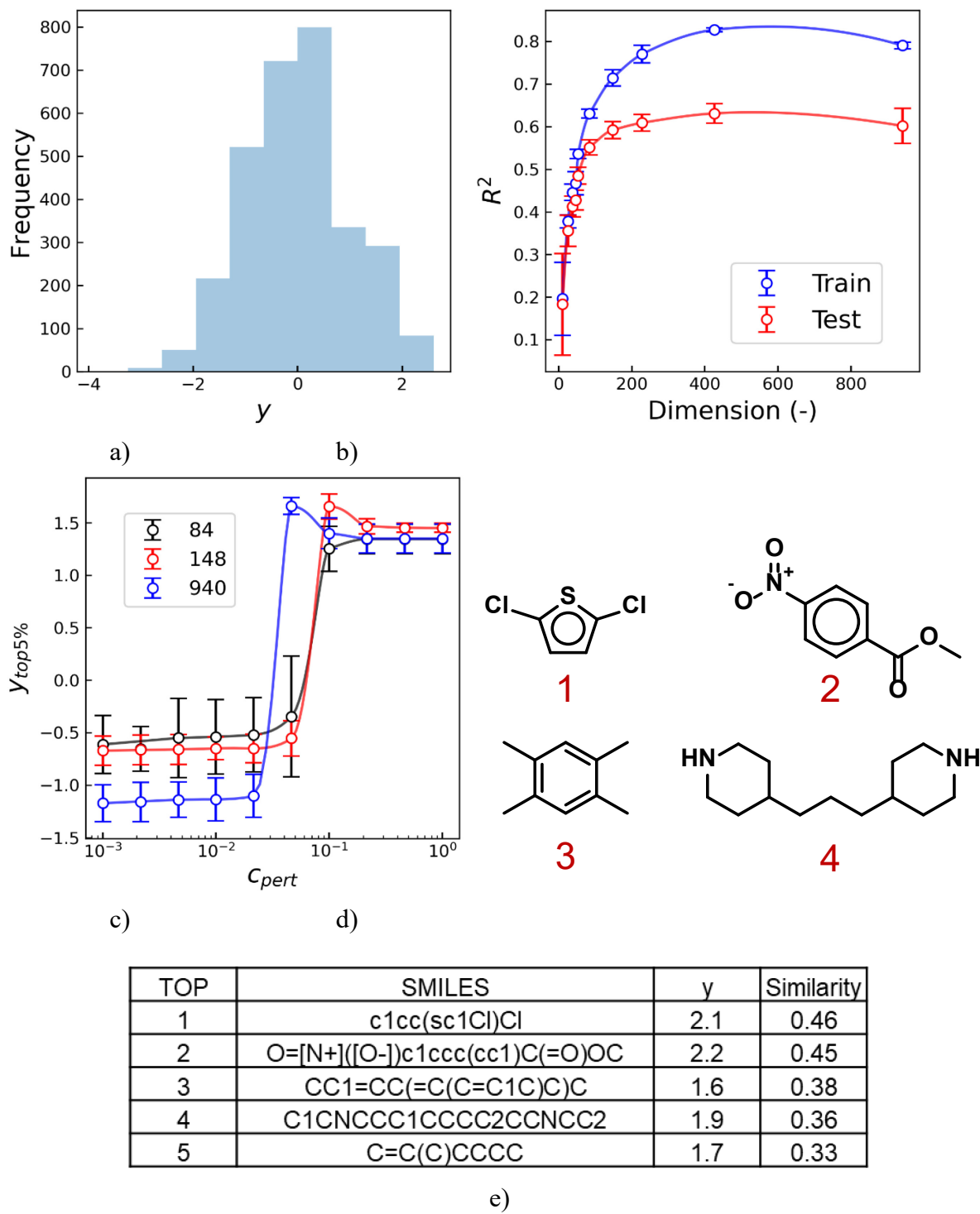

**Figure S7.** Exploration results of low-melting-temperature compounds by DAU.
a) Histogram of $y$ for the database (the sign of $y$ was changed from Figure S6). b) Regression results for the training and testing datasets. The dimension of $\boldsymbol{x}$ was changed. c) Average $y$ of the top 5% of chemicals extracted by this approach. Dimension and $c_{\text{pert}}$ were changed. Values are the averages after five-fold cross-validation. Error bars represent standard deviation. d) and e) Information on the example structures of extracted chemicals with high similarities.

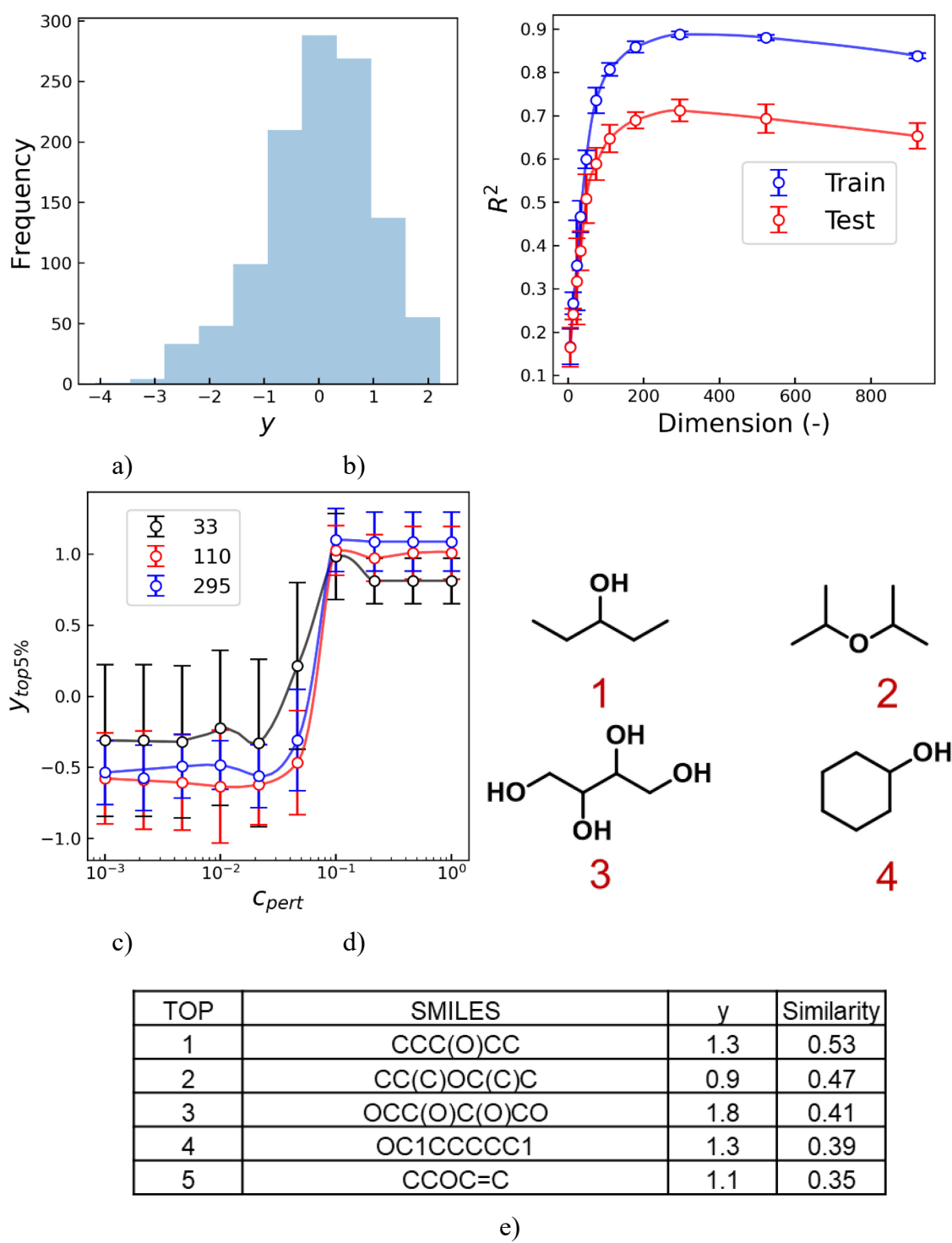

**Figure S8.** Exploration results of high-solubility-parameter compounds by DAU.
a) Histogram of $y$ for the database. b) Regression results for the training and testing datasets. The dimension of $\boldsymbol{x}$ was changed. c) Average $y$ of the top 5% of chemicals extracted by this approach. Dimension and $c_{\text{pert}}$ were changed. Values are the averages after five-fold cross-validation. Error bars represent standard deviation. d) and e) Information on the example structures of extracted chemicals with high similarities.

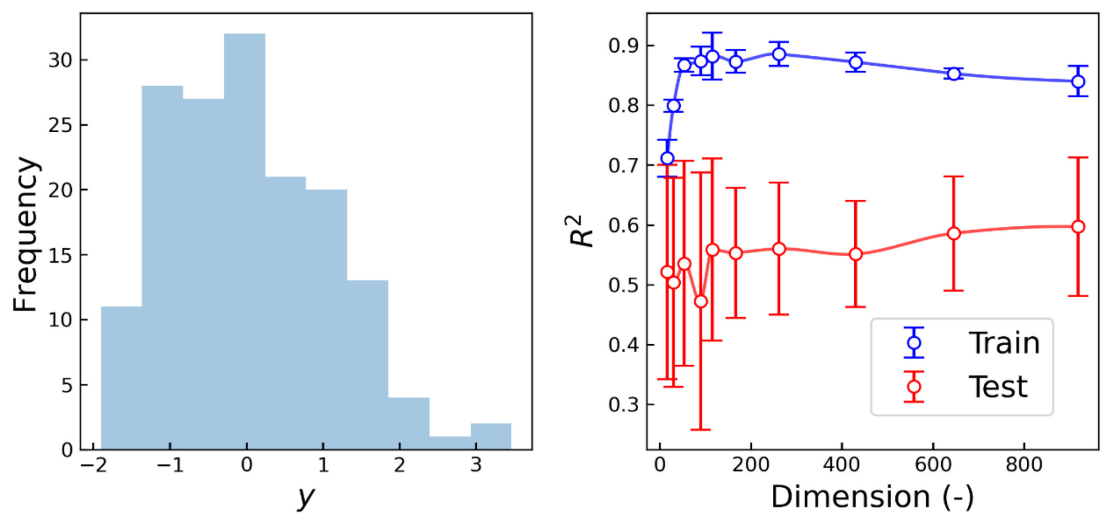

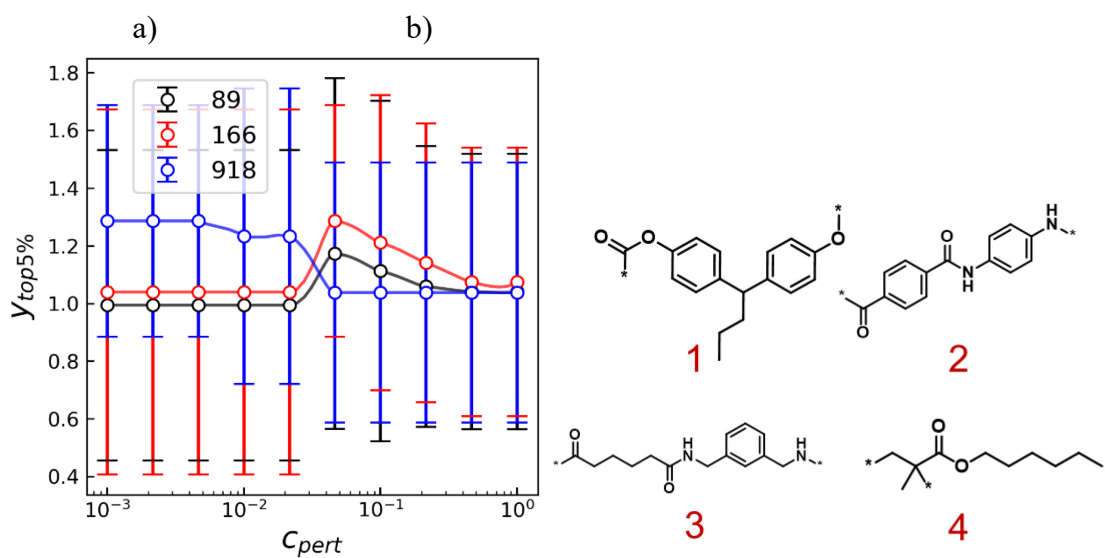

| TOP | SMILES | y | Similarity |
|---|---|---|---|
| 1 | O=C([*])OC1=CC=C(C(CCC)C2=CC=C(O[*])C=C2)C=C1 | 1.7 | 0.61 |
| 2 | [*]NC1=CC=C(NC(C2=CC=C(C=C2)C([*])=O)=O)C=C1 | 1.6 | 0.61 |
| 3 | [*]NCC1=CC=CC(CNC(CCCCC([*])=O)=O)=C1 | 1.7 | 0.58 |
| 4 | CC(C[*])([*])C(OCCCCCC)=O | 0.8 | 0.57 |
| 5 | O=C([*])OC1=CC=C(C(C2=CC=CC=C2)(C3=CC=CC=C3)C4=CC=C(O[*])C=C4)C=C1 | 1.8 | 0.52 |

e)

**Figure S9.** Exploration results of high-glass-transition-temperature polymers by DAU.
a) Histogram of $y$ for the database. b) Regression results for the training and testing datasets. The dimension of $x$ was changed. c) Average $y$ of the top 5% of chemicals extracted by this approach. Dimension and $c_{pert}$ were changed. Values are the averages after five-fold cross-validation. Error bars represent standard deviation. d) and e) Information on the example structures of extracted chemicals with high similarities.

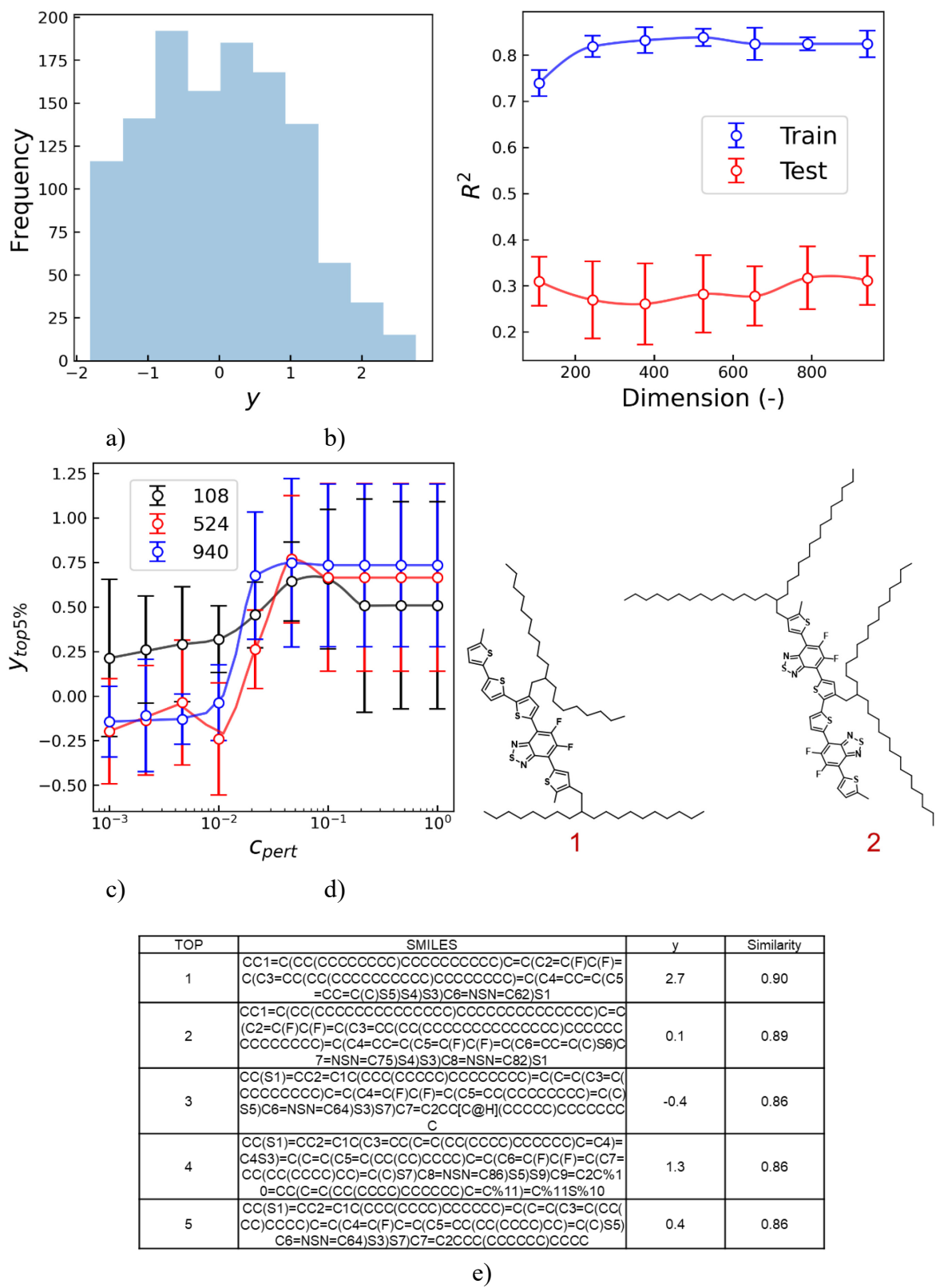

**Figure S10.** Exploration results of high-photoconversion-efficiency compounds by DAU.
a) Histogram of $y$ for the database. b) Regression results for the training and testing datasets. The dimension of $\boldsymbol{x}$ was changed. c) Average $y$ of the top 5% of chemicals extracted by this approach. Dimension and $c_{\text{pert}}$ were changed. Values are the averages after five-fold cross-validation. Error bars represent standard deviation. d) and e) Information on the example structures of extracted chemicals with high similarities.

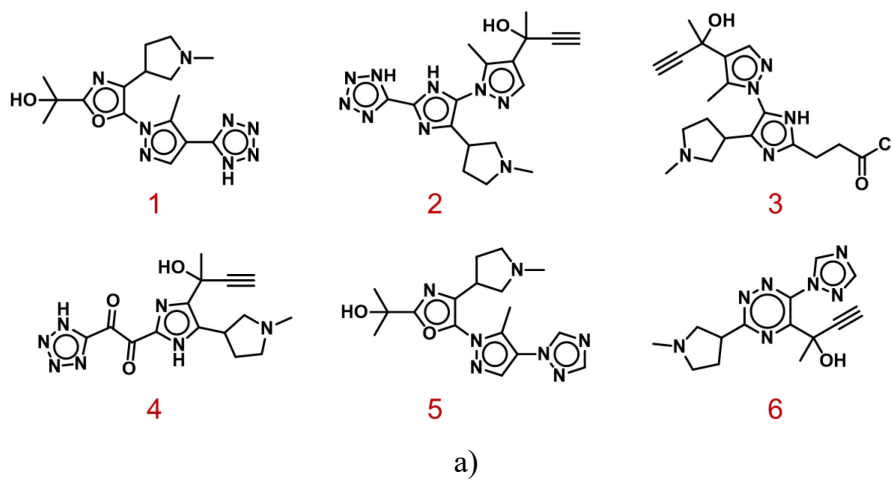

a)

| TOP | SMILES | $y_{pred}$ |
|---|---|---|
| 1 | Cc1c(-c2nnn[nH]2)cnn1-c1oc(C(C)(C)O)nc1C1CCN(C)C1 | 3.95 |
| 2 | C#CC(C)(O)c1cnn(-c2[nH]c(-c3nnn[nH]3)nc2C2CCN(C)C2)c1C | 3.89 |
| 3 | C#CC(C)(O)c1cnn(-c2[nH]c(CCC(=O)Cl)nc2C2CCN(C)C2)c1C | 3.88 |
| 4 | C#CC(C)(O)c1nc(C(=O)C(=O)c2nnn[nH]2)[nH]c1C1CCN(C)C1 | 3.86 |
| 5 | Cc1c(-n2cncn2)cnn1-c1oc(C(C)(C)O)nc1C1CCN(C)C1 | 3.83 |
| 6 | C#CC(C)(O)c1nc(C2CCN(C)C2)nnc1-n1cncn1 | 3.80 |
| 7 | C#CC(C)(O)c1nc(C2CCN(C)C2)c(-n2ncc(C(C)(O)C#C)c2C)[nH]1 | 3.77 |
| 8 | C#CC(C)(O)c1cnn(-c2[nH]c(-n3c(C)ccc3C)nc2C2CCN(C)C2)c1C | 3.73 |
| 9 | Cc1ccc(C)n1-c1[nH]c(-c2ccc(CCC(=O)Cl)c(N)c2)nc1C1CCN(C)C1 | 3.71 |
| 10 | C#CC(C)(O)c1cnn(-c2[nH]c(C(C)(C)O)nc2C2CCN(C)C2)c1C | 3.68 |

b)

**Figure S11.** a) Example structures generated from $x_{ideal}$ with high predicted melting temperatures and b) corresponding information.

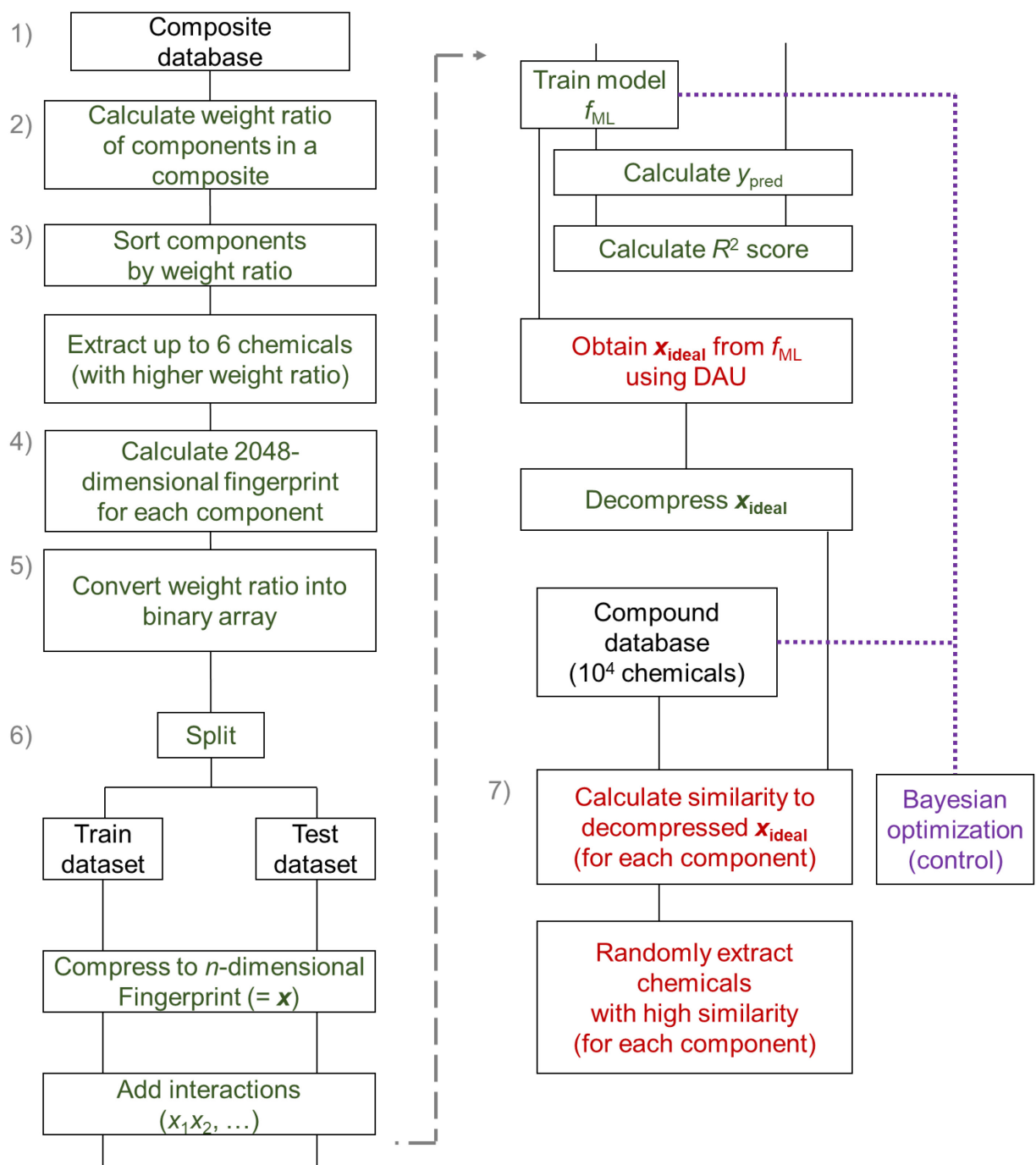

**Figure S12.** Scheme for exploring chemicals using DAU (composite system).

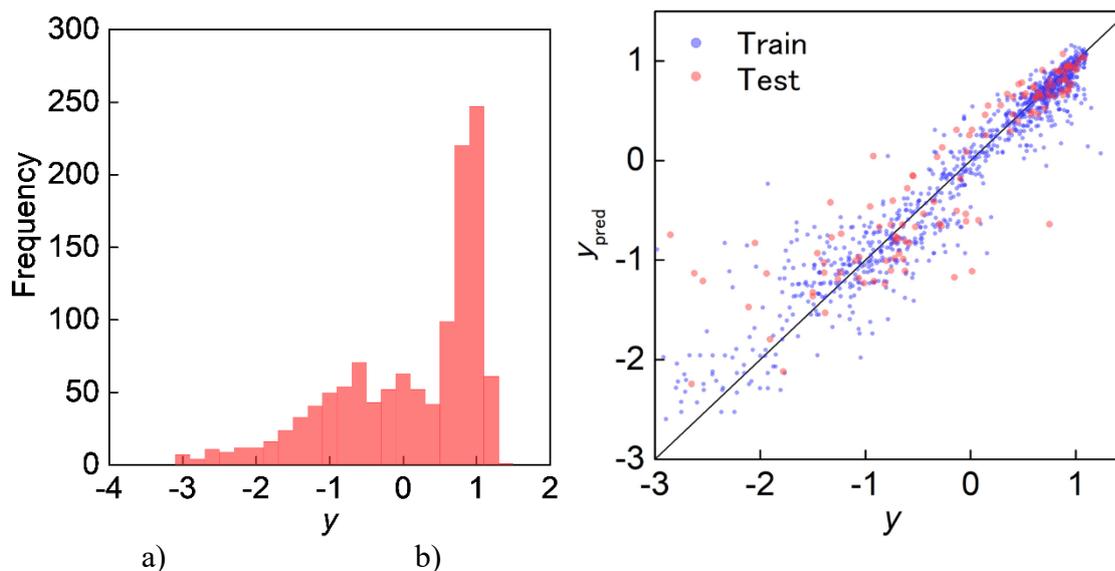

**Figure S13.** a) Histogram of $y$ for the Li$^+$-conducting electrolyte database. b) Regression results for the database. The training and testing $R^2$ scores were 0.88 and 0.68, respectively. The dimension of $x$ was 398.

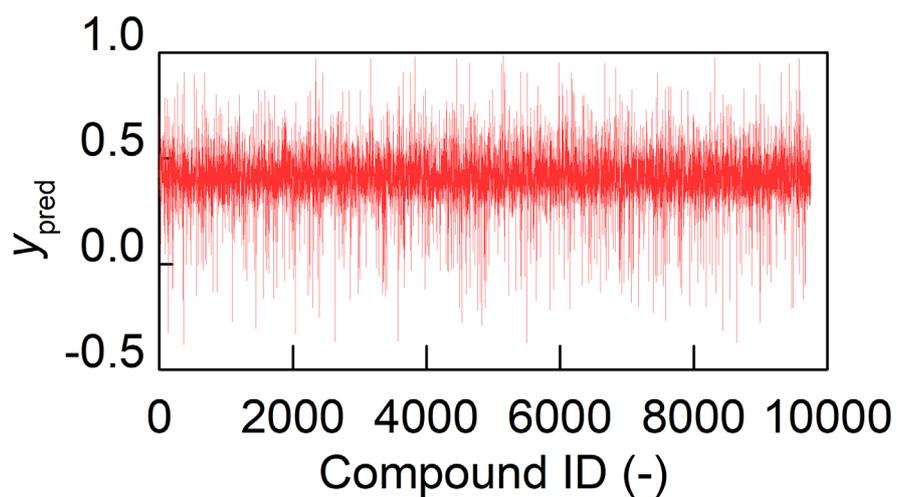

**Figure S14.** Predicted conductivity as a function of compound ID. After selecting five chemicals randomly, the sixth chemical was changed in the order of compound ID in the database. The weight ratio was also set randomly and fixed.

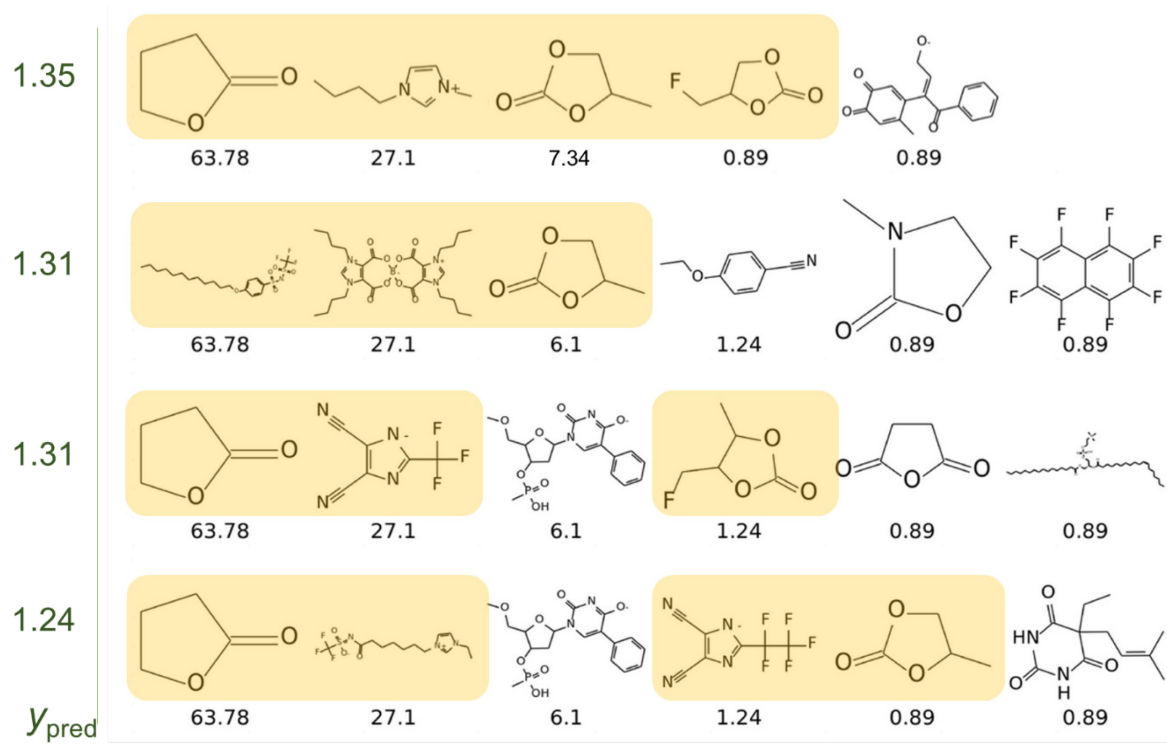

**Figure S15.** Top four compositions giving the highest $y_{pred}$, which were found by the annealing approach. The predicted $y$ of 1.35 corresponded to 47 mS/cm. Some composites are not electrically neutral, which should be changed in future work by adding restriction conditions during annealing.

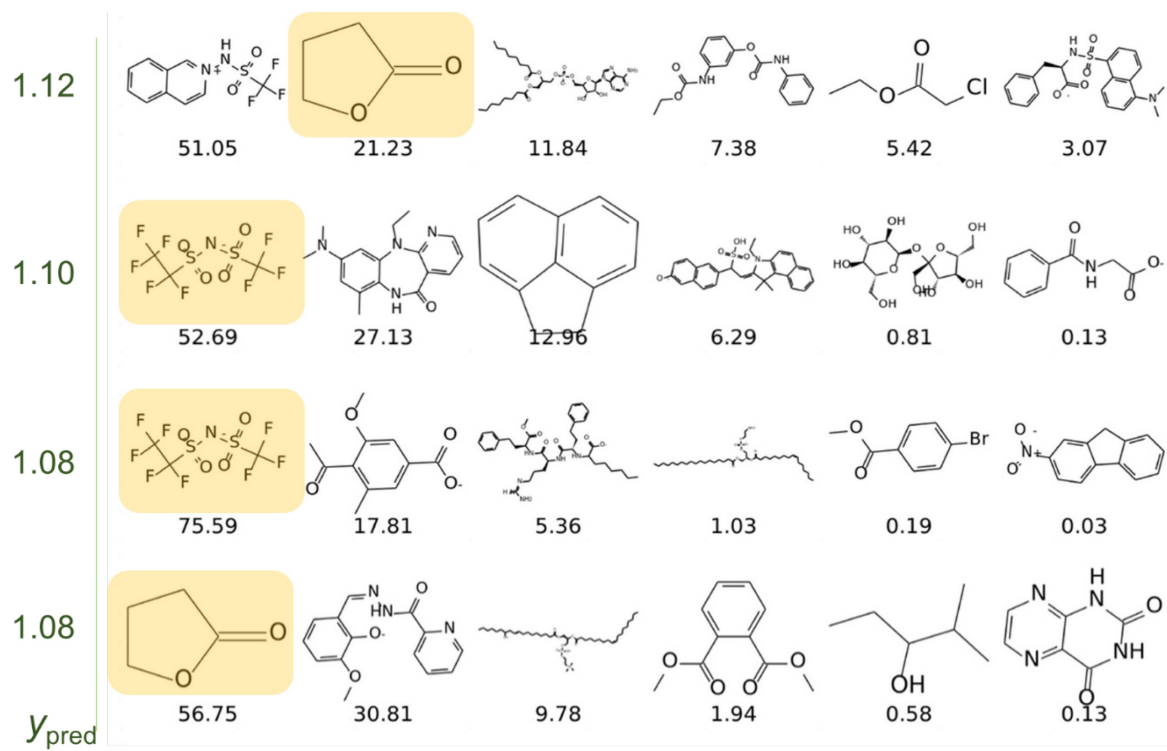

**Figure S16.** Top four compositions giving the highest $y_{pred}$, which were found by the random approach.

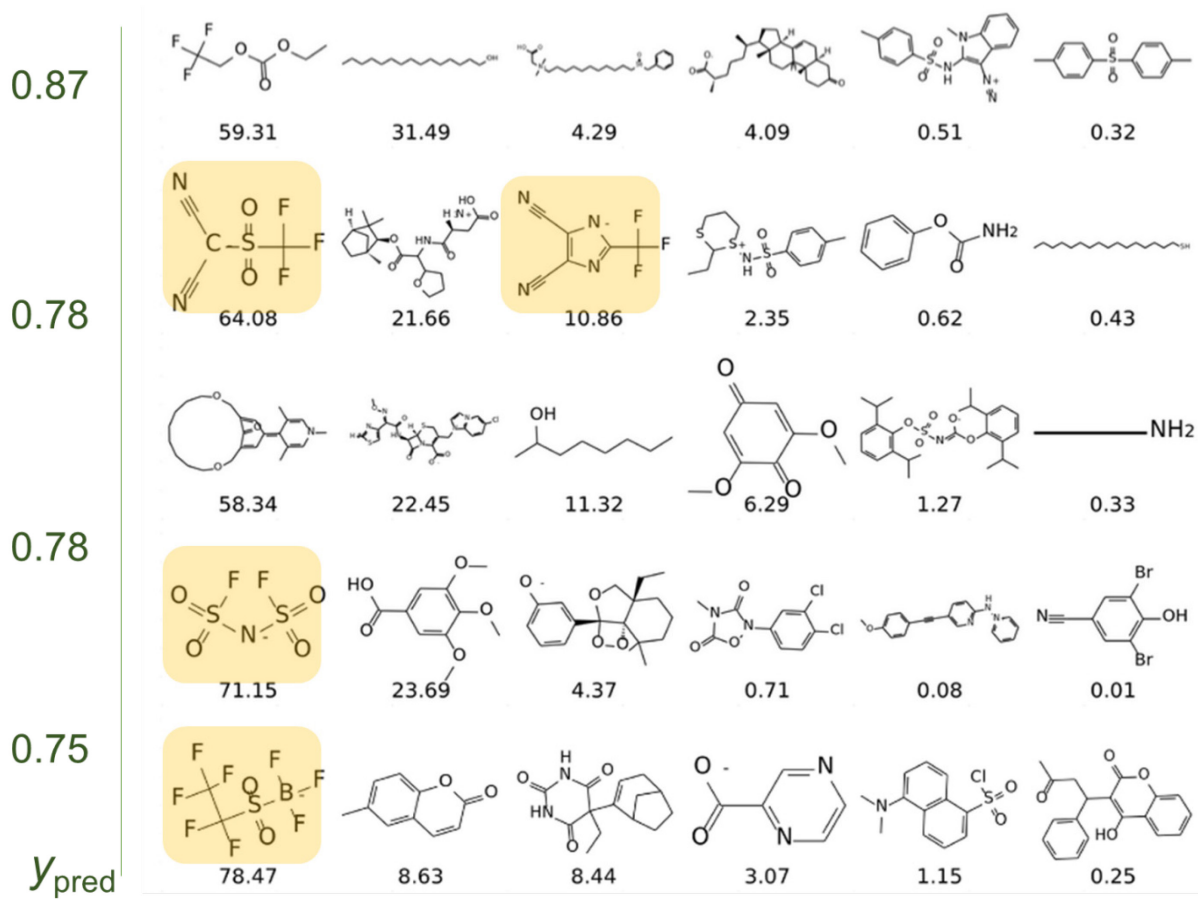

**Figure S17.** Top four compositions giving the highest $y_{pred}$, which were found by the Bayes approach.